%% file: main.tex
\documentclass[preprint,12pt,authoryear]{elsarticle}
\usepackage[margin=1.5cm]{geometry}
\usepackage[utf8]{inputenc}
\usepackage[english]{babel}
\usepackage{calc}
\usepackage{fancyhdr}
\usepackage{latexsym}
\usepackage{amsmath}
\usepackage{amssymb}
\usepackage{wasysym}
\usepackage{natbib}
\usepackage{xcolor}
\usepackage{tikz}
\usetikzlibrary{decorations.pathreplacing,automata,arrows,shadows,patterns,shapes,calc}
\usepackage{enumerate}	
\usepackage[algoruled]{algorithm2e}
\usepackage[colorlinks=true,urlcolor=blue,linkcolor=blue,plainpages=false]{hyperref}
\usepackage[all]{hypcap}
\usepackage{graphicx}
\usepackage{euscript}
\usetikzlibrary{calc}
\usepackage{comment}

\newcommand{\TODO}[1]{{\color{red}\fbox{ToDo:} {\sf #1}}}

\newcommand{\decisionpb}[3]{\fbox{\parbox{0.87\textwidth}{{#1}\\{\it Input:} #2\\{\it Question:} #3}}}

\newcommand{\inNeighbors}{\EuScript{N}^{\textrm{in}}}
\newcommand{\outNeighbors}{\EuScript{N}^{\textrm{out}}}

\newcommand{\cyclicity}{c}
\newcommand{\ssi}{\Longleftrightarrow}

\newcommand{\lcm}{\operatorname{lcm}}
\newcommand{\girth}{\operatorname{g}}
\newcommand{\landau}{\text{L}}

\newcommand{\lab}{\operatorname{lab}}
\newcommand{\mi}{\oplus}
\newcommand{\me}{\ominus}

\newcommand{\set}[1]{\left\lbrace #1\right\rbrace }
\newcommand{\paren}[1]{\left( #1\right)}

\newcommand{\refer}[3][]{\hyperref[#3]{#2~\ref*{#3}#1}}

\newtheorem{thm}{Theorem}
\newtheorem{lem}[thm]{Lemma}
\newtheorem{prop}[thm]{Proposition}
\newtheorem{cor}[thm]{Corollary}
\newproof{pf}{Proof}
\newtheorem{rem}{Remark}

\newlength{\rnodo}
\newlength{\radio}
\newlength{\Radio}
\newlength{\RADIO}
\tikzstyle{tipo}=[fill,circle,inner sep=0,text=white,minimum size=8pt]
\tikzstyle{main node}=[outer sep=1,inner sep=0,ellipse,thick,draw,minimum size=2\rnodo,fill=black!10, font=\footnotesize]
\tikzstyle{second node}=[main node, fill=black, text=white]
\tikzstyle{name node}=[outer sep=1,inner sep=0,ellipse,thick,draw,minimum size=2\rnodo,fill=black!10, font=\footnotesize]
\tikzstyle{and node}=[outer sep=1,inner sep=0,ellipse,thick,draw,minimum size=2\rnodo,fill=black!60,text=white, font=\footnotesize]
\tikzstyle{or node}=[outer sep=1,inner sep=0,ellipse,thick,draw,minimum size=2\rnodo, font=\footnotesize]
\tikzstyle{ao node}=[outer sep=1,inner sep=0,ellipse,thick,draw,minimum size=2\rnodo,fill=black!30,text=white, font=\footnotesize]
\tikzstyle{labels}=[inner sep=0pt,font=\scriptsize,auto,circle]
\tikzstyle{arcos}=[-latex,thick]
\tikzstyle{arcossym}=[latex-latex,thick]
\tikzstyle{small node}=[outer sep=1,inner sep=0,circle,thick,draw,minimum size=1\rnodo,fill=black!10]
\tikzstyle{sombra}=[outer sep=1,inner sep=0,circle,thick,draw,minimum size=1.5\rnodo, red]

\newcommand{\Loop}[4][]{\draw[arcos, #1](#2.#3+10)..controls +(#3+20:2\rnodo) and +(#3-20:2\rnodo).. node[labels]{#4}(#2.#3-10)}

\journal{Some Journal}

\begin{document}
	
	\begin{frontmatter}
		\title{Complexity of limit cycles with block-sequential update schedules in conjunctive networks}
		
		\author[conce,ciima]{Julio Aracena}
		\ead{jaracena@ing-mat.udec.cl}

		\author[mars]{Florian Bridoux\corref{cor}}
		\ead{florian.bridoux@lis-lab.fr}
		
		\author[ubb]{Luis G\'omez}
		\ead{luismiguelgomezguzman@gmail.com}

		\author[diicc,ciima]{Lilian Salinas}
		\ead{lilisalinas@udec.cl}
		\cortext[cor]{Corresponding author.}
		
		\address[conce]{Departamento de Ingenier\'ia Matem\'atica and CI$\,^2\!$MA, Universidad de Concepci\'on, Concepci\'on, Chile.}
		\address[diicc]{Departamento de Ingenier\'ia Inform\'atica y Ciencias de la Computaci\'on and CI$\,^2\!$MA, Universidad de Concepci\'on, Concepci\'on, Chile.}
		\address[ubb]{Departamento de Estadística, Facultad de Ciencias, Universidad del Bio-Bio, Concepci\'on,  Chile.}
		\address[mars]{Normandie Univ, UNICAEN, ENSICAEN, CNRS, GREYC, 14000 Caen, France.}
		
		\begin{abstract}
			In this paper, we deal the following decision problem:
			given a conjunctive Boolean network defined by its interaction digraph, does it have a limit cycle of a given length $k$?
			We prove that this problem is NP-complete in general if $k$ is a parameter of the problem and in P if the interaction digraph is strongly connected.
			The case where $k$ is a constant, but the interaction digraph is not strongly connected remains open.
			
			Furthermore, we study the variation of the decision problem:
			given a conjunctive Boolean network, does there exist a block-sequential (resp. sequential) update schedule such that there exists a limit cycle of length $k$?
			We prove that this problem is NP-complete for any constant $k \geq 2$.
		\end{abstract}
		\begin{keyword}
			Boolean network, limit cycle, update schedule, update digraph, NP-Hardness.
		\end{keyword}
		
	\end{frontmatter}
	
	\section{Introduction}

A Boolean network with $n$ components is a discrete and finite dynamical system whose dynamics can be described by a map from $\{0,1\}^n$ to $\{0,1\}^n$. Boolean networks have many applications. In particular, they are classical models for the dynamics of gene networks~\cite{kauffman1969metabolic,thomas1973boolean,thomas1990biological,thomas2001multistationarity,de2002modeling}, neural networks~\cite{mcculloch1943logical,hopfield1982neural,goles1985dynamics,goles2013neural} and social interactions~\cite{poljak1983periodical,poindron2021general}. They are also essential tools in information
theory, for the binary network coding problem~\cite{riis2007information,gadouleau2011graph,gadouleau2016reduction}.

A conjunctive network is a particular type or Boolean network where the local function of each component (the function that behaves the evolution of the state of the component) realizes a conjunction on the state of a subset of components of the network.
The study of conjunctive networks in the context of gene networks has captured special interest in the last time due to
increasing evidence that the synergistic regulation of a gene by several transcription factors, corresponding to a conjunctive function, is a common mechanism in regulatory networks~\cite{nguyen2006deciphering,gummow2006reciprocal}. Besides, they have been used in the combinatorial influence of deregulated gene sets on disease phenotype classification~\cite{park2010inference} and in the construction of synthetic gene networks~\cite{shis2013library}.
Also conjunctive networks have been studied with an analytic point of view~\cite{aracena2017fixed,GolHer,DBLP:conf/automata/Gadouleau21,Math12,jarrah2010dynamics,aledo2012parallel,gao2018stability}.


The update schedule in a Boolean network, that is the order in which each node is updated, is of great importance in its dynamical behavior. In general, Boolean networks are usually studied with synchronous (parallel) or sequential schemes. A generalization of these schemes, known as block-sequential update schedules, was introduced by F. Robert  \cite{FRob86,robert1995systemes}, and they are currently used in the modeling of regulatory networks \cite{goles2013deconstruction,ruz2014neutral}.

Many analytic studies have been done about the limit cycles of a Boolean network with different block-sequential update schedules \cite{LG13,Ara09,DBLP:conf/sofsem/BridouxGPS21,demongeot2008robustness,Math12,goles2014computational,PhDthesisLuis,macauley2009cycle,mortveit2012limit}. Most of them show that the limit cycles are very sensitive to changes in the updating scheme of a network. In particular, some of these articles exhibit examples of Boolean networks where the existence of limit cycles depends on the used update schedule \cite{LG13,Ara09,aracena2013number,DBLP:conf/sofsem/BridouxGPS21,demongeot2008robustness,mortveit2012limit}.
Some papers study the complexity of problems related to Boolean networks and their limit-cycles~\cite{DBLP:conf/sofsem/BridouxGPS21,PhDthesisLuis,LG13}.

In this paper, we study the algorithmic complexity of  problems of existence of limit cycles in conjunctive networks, highlighting the differences between the parallel schedule and other block-sequential update schedules. 

This paper is organized as follows. In sections~\ref{section:PLCE}, we study the problem of knowing if a given conjunctive network updated in parallel has a limit cycle of length $k$.
Theorem~\ref{thm:PLCE_s_P} states that the problem is in P if the interaction digraph of the conjunctive network is strongly connected.
Next, we give two side results.
Corollary~\ref{cor:prime} states that the length of the limit cycles of a conjunctive network of size $n$ has no prime factor strictly greater than $n$.
Corollary~\ref{cor:landau} states that the bigger length of a limit cycle that a conjunctive network of size $n$ can have is $\landau(n)$ where $\landau$ is the Landau's function, that is the largest $\lcm$ of numbers $t_1,\dots,t_m$ whose sum is $n$.
Theorem~\ref{thm:SLCE-NP} states that the problem is NP-complete if $k$ is a parameter of the problem given in binary.
Finally, Proposition~\ref{prop:power_of_prime} states that the problem is in $P$ when $k$ is a constant and a power of a prime.
The question of the complexity class of the problem remains open when $k$ is a constant but not a power of a prime and the interaction digraph is not strongly connected.

In section~\ref{section:BLCE}, we study the problem of knowing if, given a conjunctive network, does there exist a block-sequential (\textit{resp.} sequential) update schedule $s$ such that $f^s$ ($f$ updated in the order given by $s$) has a limit cycle of length $k$?
Theorem~\ref{thm:impl} states that if $f$ has a block-sequential update schedule $s$ such that $f^s$ has a limit cycle of length $k>2$ then $f^s$ has a sequential (and therefore block-sequential) update schedule such that $f^s$ has a limit cycle of length $k-1$.
Theorem~\ref{thm:NP_c} states that the problem is NP-complete of all $k \geq 2$ for the two versions of the problem: block-sequential and sequential.

\section{Definitions and notation}\label{sec:def}

%

A \emph{Boolean network} (BN) $N$ of size $n$ can be represented as a (global) function $f: \set{0,1}^n \to \set{0,1}^n$ and an update schedule $s$.
The global function $f:\set{0,1}^n \to \set{0,1}^n$ can be decomposed into $n$ \emph{local functions} $f_1, \dots, f_n: \set{0,1}^n \to \set{0,1}$, each local function describing the behavior of a component of the BN and for all \emph{configurations} $x \in \set{0,1}^n$, we have $f(x) = (f_1(x), \dots f_n(x))$.
The update schedule defines the order in which the components of the BN are updated between two time steps.
This paper is focused on the \emph{block-sequential update schedules} which were introduced in~\cite{FRob86}:
the coordinates $[n] := \set{1,\dots,n}$ are partitioned into $p$ blocks $s=(B_1, \dots, B_p)$ and the dynamics of $N = (f,s)$ is defined by:
\[
f^s = f^{B_p} \circ \dots \circ f^{B_1}
\]
with $f^{B_i}:\{0,1\}^n\to \{0,1\}^n $ such that:
\[
f_j^{B_i}(x) =\begin{cases}
    f_j(x) & \text{if } j\in B_i\\
    x_j & \text{if } j\notin B_i\\
\end{cases}
\]
The dynamics of a BN $N = (f,s)$ is given by the function $f^s$ and two Boolean networks $N_1=(f,s)$ and $N_2=(f',s')$ have the same dynamical behavior if $f^{s}=(f')^{s'}$.
There are two extreme cases of block-sequential update schedules: 
\begin{itemize}
	\item Parallel update schedule (denoted $s^p$): there is only one block $[n]$.
	In other words, all components are updated all together and the dynamics of $N = (f,s^p)$ is simply $f$.
	\item Sequential update schedule: all components are updated one at the time.
	The dynamics of $N = (f,s)$ is then given by $f^s = f^{s_n} \circ \dots \circ f^{s_1}$ (with $f^{ i } := f^{ \{ i \} }$ for all $i \in [n]$).
\end{itemize}

Note that the BN $N = (f,s)$ with a block-sequential update schedule has the same dynamics $f^s$ that the BN $N' = (f^s,s^p)$.
In other words, all dynamics can be realized with a parallel update schedule.
Also, the block-sequential update schedule was called Serial-Parallel in~\cite{FRob86}, and in the particular case of sequential updates, $f^s$ was called Gauss-Seidel operator.
In the following, for any $i \in [n],$ $s(i) \in [p]$ is the number of the block in which $i$ is updated.

Since $\set{0,1}^n$ is a finite set, for any BN $f:\set{0,1}^n \to \set{0,1}^n$ and update schedule $s$,
we have two limit behaviors for the iteration of a network:
\begin{itemize}
	\item \emph{Fixed Point}. A fixed point is a configuration $x\in \set{0,1}^n$ such that $f^s(x)=x$.
	\item \emph{Limit Cycle}. A limit cycle of length $\ell>1$ is a tuple $(x^0,x^1,x^2, \dots,x^{\ell-1})$ such that
	\begin{itemize} 
		\item for all $i,j \in [0,\ell[$, if $i \neq j$ then $x^i \neq x^j$, and 
		\item for all $i \in [0,\ell[$, $x^{(i+1) \bmod \ell} = f^s(x^i)$.
	\end{itemize}
\end{itemize}

Fixed points and configurations in limit cycles are called \emph{attractors} of the network.
In the following, we use these notations: given a BN of dynamics $f$ and $k \geq 2$,
\begin{itemize}
	\item $\Phi_k(f) :$ set of configurations in a limit cycle of length $k$ of $f$.
\end{itemize}

We also note: $\Phi(f) = \bigcup_{k \geq 2} \Phi_k(f)$.

For any attractor $x \in \Phi_k(f)$ and for any $u \in [n]$, the periodic trace of $u$ is the word $\rho_u(x) = (x_u,f_u(x),f^2_u(x), \dots, f^{\ell-1}_u(x))$ with $\ell \in [1,k]$ the minimum number such that for all $t \in \mathbb{N}$, $f^{t+\ell}_{u}(x) = f^{t}_u(x)$.

\subsection{Interaction digraphs}

A digraph $D = (V,A)$ is composed of a set of vertices $V$ and a set of arcs $A \subseteq V \times V$.
An arc $(v,v)\in A$ is called a \emph{loop}. 
Given a vertex $v\in V$, the set of its incoming neighbors is denoted as
\mbox{$\inNeighbors_D \paren{v}=\set{u\in V \colon \paren{u,v}\in A}$}. 
Analogously, the set of outgoing vertices from $v$ is denoted as \mbox{$\outNeighbors_{D}\paren{v}=\set{u\in V\colon \paren{v,u}\in A}$}.
A strongly connected component is \emph{not trivial} if it has at least one arc.

Given a non-trivial strongly connected component $H$ of a digraph $D$, the \emph{index of cyclicity} of $H$, denoted $\cyclicity(H)$, is defined as the greatest common divisor of the lengths of the cycles of $H$.
If a digraph $D$ has non-trivial strongly connected components $H_1, \dots H_m$, then its index of cyclicity is defined as $\cyclicity(D)=\lcm\{\cyclicity(H_i): 1\leq i\leq m\}$ (with $\lcm$ the \emph{least common multiple}), or one if it does not have any cycles.
The index of cyclicity was referred to as the \emph{loop number} in \cite{AndOr08} where it was proved that it can be computed on polynomial time in \cite{colon2005boolean}.
\begin{lem} [\cite{colon2005boolean}] \label{lem:poly_cicli}
	The index of cyclicity $\cyclicity(H)$ of a non-trivial strongly connected component $H$ of a digraph $D$ can be computed in polynomial time.
\end{lem}

The number $\cyclicity(D)$ is also referred to as the \emph{index of cyclicity} of $D$ in~\cite{InCy} and as the \emph{index of imprimitivity} of the adjacency matrix of $D$ in~\cite{brualdi1991combinatorial} and \cite{InPr2}. 
We say that $D$ is \emph{primitive} if $\cyclicity(D)=1$. 
The \emph{girth} of a digraph $D$ (\textit{resp.} a component $H$), denoted $\girth(D)$ (\textit{resp.} $\girth(H)$), is defined as the length of the shortest cycle in $D$ (\textit{resp.} in $\girth(H)$).
Naturally, we have $\cyclicity(H) \leq \girth(H)$. 

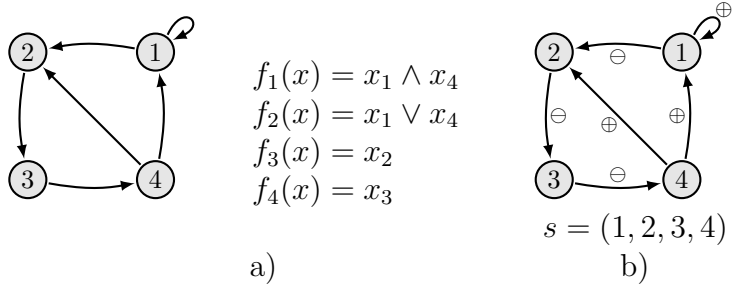
\begin{figure}[h]
	\centering
	\input{fig/FigEjID}
	\caption[Example of an interaction digraph and an update digraph.]
	{a) Digraph associated to a global function $f$. b) Update Digraph associated to a Boolean network $N = (f,s)$.}\label{fig:AssDig}
\end{figure}

The \emph{interaction digraph} of a global function $f: \{0,1\}^n \to \{0,1\}^n$, is the directed digraph $D^f=(V,A)$, where $V = [n]$ and $(u,v)\in A$ if and only if $f_{v}(x)$ depends on $x_{u}$.
More formally, $(u,v)\in A$ iff there exists \mbox{$x\in\set{0,1}^n$} such that $f_{v}(x)\neq f_{v}(\bar{x}^{u})$ (where $\bar{x}^{u} \in \{0,1\}^n$ is a configuration different of $x$ in the coordinate $u$ and identical everywhere else).
Note that if $f_{v}$ is constant, then $\inNeighbors_{D^{f}}\paren{j}=\emptyset$. See an example of a interaction digraph in \refer{\figurename}{fig:AssDig}.

A \emph{$k$-labeling} function $\lab$ of a digraph $D=(V,A)$ is a labeling function of $D$ such that all cycles of $D$ have a multiple of $k$ positive arcs.
The labeled digraph $D_{\lab}$ is then called a \emph{$k$-labeled} digraph.

\begin{rem}
	For any integer $k$ and any digraph $D=(V,A)$, there is always a $k$-labeling function: 
	the function $\lab:A \mapsto \me$ that associates a negative sign to all arcs of $D$.
	However, if it is not an acyclic digraph, this function does not correspond to an update digraph.
\end{rem}

Consider a digraph $D=(V,A)$ of size $n$ and a block-sequential update schedule $s=(B_1, \dots, B_p)$.
We denote \mbox{$D_s=(D,\lab_s)$} the labeled digraph, called \emph{update digraph}, where the function \mbox{$\lab_s:A \rightarrow \{ \me,\mi \}$}  is defined as: $\forall (u,v) \in A, \ u \in B_i \land v \in B_j$:

\begin{equation}\label{def:UD}
\lab_s(u,v)=\begin{cases}
\mi &\text{ if } i \geq j, \\
\me &\text{ if } i <j.
\end{cases}
\end{equation}

If $s$ is a sequential update schedule, $D_s$ is then called a \emph{sequential update digraph}.

The update digraph associated to a Boolean network $N=(f,s)$ is defined by \mbox{$D^f_s =(D^f, \lab _s)$} (see an example of update digraph $D_s$ in~\refer{\figurename}{fig:AssDig}). 


Note that the label on a loop will always be $\mi$. 
It was proven in~\cite{Ara09} that if two Boolean networks with the same update function and different updates schedules give the same update digraph, then they also have the same dynamical behavior (\textit{i.e.} $D^f_s = D^f_{s'} \implies f^{s} = f^{s'}$).

Given an update digraph $D_s$, with $D=(V,A)$, we define the operator $\mathcal{P}$ as \mbox{$\mathcal{P}(D_s)=(V,A')$}, where $(u,v)\in A'$ if and only if
there exists a path $(u_1,u_2, \dots,u_m)$ with $u = u_1$ and $v = u_m$ in $D$ such that $\lab(u_1=u,u_2)=\mi$ and $\lab(u_2,u_3)= \lab(u_3,u_4) = \dots = \me$  (see also \cite{Math12}). See an example in \refer{\figurename}{fig:PDig}. $\mathcal{P}(D_s)$ is referred to as the \emph{parallel digraph of $D_s$}. 

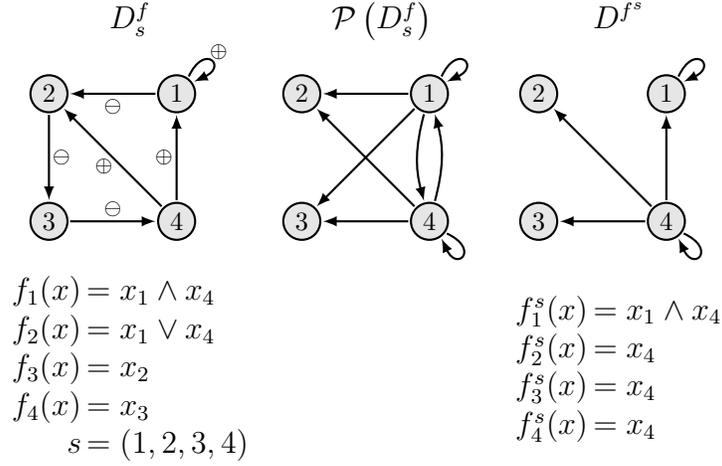
\begin{figure}[h]
	\centering
	\input{fig/PGFs}
	\caption{Example of $D_s^f$, $\mathcal{P}(D_s^f)$ and $D^{f^s}$}\label{fig:PDig}
\end{figure}


%


A local function $f_v: \{0,1\}^n\to\{0,1\}$ is said \emph{conjunctive} if there exists a set $I \subseteq [n]$ such that \mbox{$f_v(x)=0\ssi\exists u\in I$ such that $x_u=0$} (if $I = \emptyset$ then $f_v(x) = 1$ for any $x \in \{0,1\}^n$). 
A {conjunctive global function} $f\colon\set{0,1}^{n}\to\set{0,1}^{n}$ is a global function where each local function is conjunctive. A conjunctive global function $f$ can be completely described by its interaction digraph $D^f$.
That is, given a digraph $D=\paren{V,A}$ , we have for every $v \in V$ :
\[
f_v(x) = \bigwedge\limits_{u \in \inNeighbors_D(v)} x_u.
\]
Note that if $\inNeighbors_{D}\paren{v}=\emptyset$, then $f_v\paren{x}= 1$.

\begin{figure}[htb]
	\centering
	\input{fig/FigPrimitive}
	\caption{Example of (from left to right): two updates digraphs, their associated parallel digraphs, and their dynamical behavior.}\label{fig:Primitive}
\end{figure}
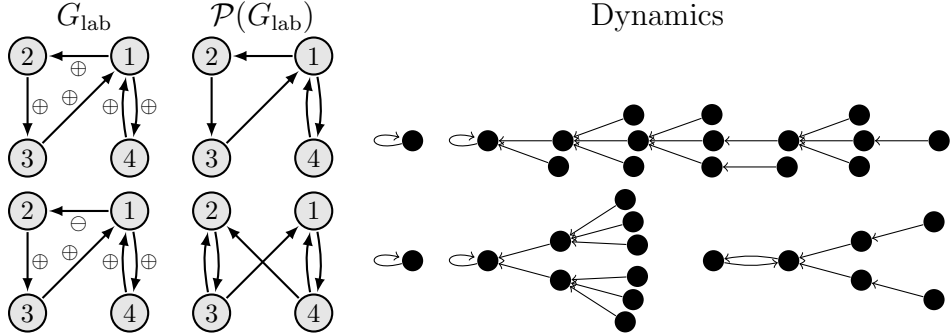

\subsection{3-SAT}

A 3-CNF formula $\psi$ is composed of $n$ variables $\lambda = \{ \lambda_1, \dots, \lambda_n \}$ and $m$ clauses $\mu = \{ \mu_1, \dots, \mu_m \}$. 
Each clause $\mu_j \in \mu$ is a set of three literals $ \{ \mu_{j,1},\mu_{j,2},\mu_{j,3} \}$. 
A literal $\mu_{j,p}$ is a couple composed of a variable $\lambda_i \in \lambda$ and a polarity $\rho \in \{ \top, \bot \}$.
A clause $\mu_j \in \mu$ is satisfied by a valuation $g: \lambda \to \{ \bot, \top \}$ if 
there is a literal $\mu_{j,p} = ( \lambda_i,\rho) \in \mu_j$ such that $g(\lambda_i) = \rho$. 
By abuse of notation, we then write $g(\mu_{j,p}) = \top$ (the valuation satisfies the literal) and $g(\mu_{j}) = \top$ (the valuation satisfies the clause).
The decision problem 3-SAT is to know if, given a 3-CNF formula $\psi$,
there exists a valuation $g: \lambda \to \{ \bot, \top \}$ which satisfies $\psi$ (\textit{i.e.} each clause of $\psi$).
By Cook's theorem, the problem 3-SAT is NP-hard (see Theorem 8.2 of~\cite{P94}).

\section{Decision problems}

In this section, we focus on conjunctive networks and define the following three decision problems  for any constant $k$. 

\begin{quote}
	\decisionpb
	{\sc Parallel Limit Cycle Existence ($k$-PLCE) Problem:} 
	{A global function $f:\{0,1\}^n \to \{0,1\}^n$.}
	{Does $\Phi_k(f) \neq \emptyset$?}
\end{quote}

\begin{quote}
	\decisionpb
	{\sc Block-sequential Limit Cycle Existence ($k$-BLCE) Problem:} 
	{A global function $f:\{0,1\}^n \to \{0,1\}^n$.}
	{Does there exist a \textbf{block-sequential} update schedule $s$ such that $\Phi_k(f) \neq \emptyset$ ?}
\end{quote}

\begin{quote}
	\decisionpb
	{\sc Sequential Limit Cycle Existence ($k$-SLCE) Problem:}
	{A global function $f:\{0,1\}^n \to \{0,1\}^n$.}
	{Does there exist a \textbf{sequential} update schedule $s$ such that $\Phi_k(f^s) \neq \emptyset$?}
\end{quote}
We will also study a variant of these three decision problems where $k$ is not part of the problem, but a parameter of the problem encoded in binary.
These three new problems are then simply named PLCE, BLCE and SLCE.

In general, for these problems, $f$ has to be encoded efficiently because  there are $(2^n)^{(2^n)}$ distinct global functions from $\{0,1\}^n$ to $\{0,1\}^n$.
A possible solution is to only consider global functions with an interaction digraph with bounded incoming degree. 
In the following, we consider these problems restricted to \textbf{conjunctive} global function $f$.
This type of function can be represented efficiently by its interaction digraph $D^f$ and $D^f$ itself can be represented by an adjacency matrix encoded in $n^2$ bits.

\begin{rem}
 The configuration $(1)^n$ is always a fixed point of a conjunctive function $f$ and therefore the cases $k=1$ is trivial and not interesting.
\end{rem} 

A digraph $D$ is \emph{cyclically $k$-partite} if its vertex set can be partitioned into $k$ parts $V_0, \dots , V_{k-1}$ such that the graph obtained by collapsing each part into a single vertex is a cycle of length $k$, that is,
every arc of $D$ goes from $V_i$ to $V_{(i+1) \bmod k}$ for some $0 \leq i \leq k-1$. 
The same definition can be used to a non-trivial strongly connected component of a digraph $D$. 

There is a direct relation between the index of cyclicity and this notion of ``cyclically $k$-partite".
\begin{lem}[{\cite{brualdi1991combinatorial}}]  \label{lem:partite} 
	If $D$ is a strongly connected digraph with index of cyclicity $\cyclicity(D) = k$, then D is cyclically $k$-partite.
\end{lem}

Furthermore, there is a relation between the index of cyclicity and the length of the limit cycles.
\begin{lem}[{\cite{GolHer,AndOr08}}] \label{lem:plce}
	Let $D$ with a unique strongly connected component with index of cyclicity $\ell$, and let $f$ be the conjunctive
	network on $D$. Then, $\Phi_k(f) \neq \emptyset$ if and only if $k$ divide $\ell$.
\end{lem}

Consequently, if the index of cyclicity is $1$ then $f$ has only fixed point as attractors. 

Conjunctive networks updated under another kind of update schedules were studied in \cite{Math12}, where, in our wording, was shown the following result:
\begin{prop}\label{LCEORsym}
	Let $f$ be a conjunctive global function with symmetric interaction digraph $D^f$. 
	Then, if $D^f$ is strongly connected, for all $s\neq s^p$ such that $D^f_{s} \neq D^f_{s^p}$, $\Phi_{\geq 2}(f^s)=\emptyset$.
	Furthermore, all limit cycles of $f$ are of length 2 if $D^f$ is bipartite and $\Phi_{\geq 2}(f) = \emptyset$ otherwise.
\end{prop}

Observe that, if $f$ is a conjunctive global function with $D^f$ a complete digraph, then for every update schedule $s$, $\Phi_{\geq 2}(f^s)=\emptyset$.

It is clear that the condition in the above proposition can be tested in polynomial time, and therefore the problems with symmetric interaction digraphs are polynomial. We are now considering the problem in the general case.

In \refer{Lemma}{lem:plce}, it is claimed that the dynamics of the limit cycles of conjunctive networks with strongly connected interaction digraph updated in parallel is polynomial characterized.
Hence, our approach will consist in studying: 
\begin{itemize}
\item conjunctive global function $f$ with not strongly connected digraph $D^f$, and
\item conjunctive global function $f$ updated with other block-sequential update schedules $s$ by studying $D^{f^s}$.
\end{itemize}

Constructing the interaction digraph from the global function $f^s$ is not an easy task in general, since several compositions of conjunctive functions must be made. However, it is enough to work with the most easily to construct $\mathcal{P}(D_s^f)$, since next lemma shows that they are both the same digraph.

\begin{lem} [{\cite{Math12}}]\label{lem:effectivearcs}
	If $f$ is a conjunctive global function, then $f^s$ is also a conjunctive global function and $D^{f^s}=\mathcal{P}(D_s^f)$.
\end{lem}

\begin{pf} Straightforward from the definition of operator $\mathcal{P}$ and the fact that the composition of conjunctive functions is also a conjunctive function (and therefore $f^s$ is a conjunctive function).
	\qed
\end{pf}





\section{Complexity of the PLCE and $k$-PLCE problems} \label{section:PLCE}

In this section we study the problem of knowing if a given conjunctive network $f$ with a parallel update schedule has a limit cycle of length $k$. 

We define the circular right shift $\sigma$ as the function that take a word $a = (a_1, \dots, a_m)$ on the alphabet $\{0,1\}$ and return a word $b \in \{0,1\}^{m}$ where $b_i = a_{i-1}$ for all $2\leq i \leq m$ and $b_1 = a_m$.
Furthermore, for all $t \in \mathbb{N}$, $\sigma^t$ is the composition of $t$ times the function $\sigma$.

Proposition~\ref{prop:cycle_in_orbit} below gives some proprieties on the behavior of the cycle $c$ of a configuration $x$ in a limit cycles.

\begin{prop} \label{prop:cycle_in_orbit}
	Let $f$ be a conjunctive global function, and let $x \in \Phi_k(f)$.
	Then, for any cycle $c = (u_1,\dots,u_{m})$ of $D$ we have:
	\begin{enumerate}
	    \item \label{item:c1} For all $t\geq 0$, $| \{ u_i \in c : f^t(x)_{u_i} = 0 \} | = b$ with $b$ a constant.
	    \item \label{item:c2} For all $t\geq 0$, $f^t_c(x) = \sigma^t(x_c)$.
	    \item \label{item:c3} $f_c^{m}(x) = x_c$.
	\end{enumerate}
\end{prop}
\begin{pf}
	Remark that for any configuration $y \in \{0,1\}^n$, the function $f$ can only increase the number of $0$ present in the cycle $c$.
	Indeed, for any $u_i \in c$ such that $x_{u_i} = 0$, 
	\[
	f_{u_{i+1}}(x) = \bigwedge_{v \in \inNeighbors(u_{i+1})} x_v = x_{u_i} \wedge \dots = 0 \wedge \dots  = 0
	\]
	(where if $i = m$, $i+1 = 1$).
	If $f_c(x)$ has more $0$ than $x_c$ then $f^k_c(x) = f^{k-1}_c(f(x))$ has also more $0$ than $x_c$ and $f^k(x) \neq x$.
	It would be a contradiction because $x \in \Phi_k(f)$. 
	So the number of 0 present in the cycle $c$ in a configuration $x \in \Phi_k(f)$ is invariant by $f$ and this proves item~$\ref{item:c1}$.
	
	Now, since by item~$\ref{item:c1}$, the number of $0$ in $x_c$ is the same as in $f_c(x)$, the vertices $u_i$ such that $f_{u_i}(x) = 0$ are exactly those that $x_{u_{i -1}}=0$ (again, with $i-1=m$ if $i=1$).
	This means that if $x_{u_{i-1}} = 1$ then $f_{u_i}(x) = 1$ and therefore  $f_{u_{i}}(x) = x_{u_{i-1}}$ and $f_{(u_1,u_2,\dots,u_{m})}(x) = x_{(u_{m},u_1,u_2,\dots,u_{m-1})} = \sigma(x_c)$.
	
	Lastly, since the function $f$ shifts circularly the cycle $c$ ($f_c(x) = \sigma(x_c)$) to the right,  by applying $m$ times the function $f$, we shift $m$ times to the right and get back to $f_c^{m}(x) = \sigma^m(x_c) = x_c$. \qed
	
\end{pf}

Proposition~\ref{prop:ntscc} below gives some proprieties on the behavior of a non-trivial strongly connected component $H$ of a configuration $x$ in a limit cycles.

\begin{prop} \label{prop:ntscc}
	Let $f$ be a conjunctive global function, let $H$ be a non-trivial strongly connected component of $D^f$ and let $x \in \Phi_k(f)$.
	Let $t_H$ be the smallest integer such that $(f^{t_H}(x) )_{H} = x_H$.
	Then we have:
	\begin{enumerate}
	    \item \label{prop:en0} For all $v \in H$, $u \in \inNeighbors_H(v)$,
	    $\rho_v(x) = \sigma(\rho_u(x))$.
	    \item \label{prop:en1} For all $v \in H$, either:
	    \begin{itemize}
	        \item For all $u \in \inNeighbors(v)$, $x_u = 1$ (and $f_v(x) = 1$), or
	        \item for $u \in \inNeighbors_H(v)$,  $x_u = 0$ (and $f_v(x) = 0$).
	    \end{itemize}
	    \item \label{prop:en2} $f_H(x)$ can be computed from only $x_H$.
	    \item \label{prop:en3} The periodic trace $\rho_H(x)$ is of length $t_H$.
	    \item \label{prop:en4} Let $u \in H$. For all $v \in H$, the periodic trace of $v$ is of length $t_H$ and $\rho_v(x) = \sigma^i( \rho_u(x))$ where $i$ is the length of any path between $u$ and $v$ modulo $t_H$.
	    \item \label{prop:en5} $D_H$ is cyclically $t_H$-partite with a partition $H_0,\dots,H_{t_H-1}$.
        \item \label{prop:en6} $ \forall i \in [0,t_H-1], $ either $x_{H_i} = 0^{|H_i|}$ or $x_{H_i} = 1^{|H_i|}$.
        \item \label{prop:en7} If $t \neq 1$ then there exists $i,j \in [0,t_H-1]$ with $x_{H_i} = 0^{|H_i|}$ and $x_{H_j} = 1^{|H_j|}$.
        \item \label{prop:en8} $t_H$ divides $k$.
        \item \label{prop:en9} $t_H$ divides $c(D_{H})$ and therefore $1 \leq t_H \leq g(H) \leq |H| \leq n$.
    \end{enumerate}
\end{prop}
\begin{pf}
    
    Let $v \in H$ and $u \in \inNeighbors_H(v)$.
    Since $H$ is a not trivial strongly connected component, there exists a path from $v$ to $u$.
    Thus, $u$ and $v$ are in a cycle $c = (c_0,c_{m-1},c_{m-2},\dots,c_{2},c_{1})$ of $H$ where $v=c_{m-1}$ and where $u=c_{0}$ is the predecessor of $v$ in $c$.
	By item~\ref{item:c2} of Proposition~\ref{prop:cycle_in_orbit}, for all $t \geq 0$,
	$f^t_u(x) = x_{c_{t \bmod m}}$ and $f^t_v(x) = x_{c_{t-1 \bmod m}}$. 
	Therefore, we have $\rho_u(x) = x_{(c_0,c_1, \dots, c_{t_c-1})}$ with $t_c$ dividing $m$ and
	$\rho_v(x) = x_{(c_1,c_2, \dots, c_{t_c-1},c_0)} = \sigma(\rho_u(x))$.
	This proves item~\ref{prop:en0}.
	
	Let $v \in H$ and $u,u' \in \inNeighbors_H(v)$.
	By item~\ref{prop:en0}, we have $\rho_v(x) = \sigma(\rho_u(x))$ and  $\rho_v(x) = \sigma(\rho_{u'}(x))$. Therefore, $\rho_u(x) = \rho_{u'}(x) = \sigma^{-1}(\rho_v(x))$ and in particular $x_{u} = x_{u'}$. 
	
    Now, let $v \in H$ and $u \in \inNeighbors_H(v)$ with $x_u = 1$. 
	By item~\ref{prop:en0}, $\rho_v(x) = \sigma(\rho_u(x))$ and in particular $f_v(x) = x_u = 1$.
	Furthermore, $f_v(x) = \bigwedge\limits_{u' \in \inNeighbors(v)} x_{u'}$. 
	So for all $u' \in \inNeighbors(v)$, $x_{u'} = 1$ and this prove item~\ref{prop:en1}.
	
	For all $v \in H$, let $u \in \inNeighbors_H(v)$, by item~\ref{prop:en1}, we have $f_v(x) = x_u$. Therefore, we can compute all $f_H(x)$ simply from $x_H$ and this proves item~\ref{prop:en2}.
	
	Since by item~\ref{prop:en2} $f_H(y)$ depends only of $y_H$ for all $y \in \Phi_k(f)$ and since $f^{t_H}_H(x) = x_H$, we can show by induction that for all $t\geq 0,$
	\[ 
	f^t_H(x) = f^{t \bmod t_H}_H(x).
	\]
	Therefore, the periodic trace $\rho_H(x)$ is of length at most $t_H$.
	Furthermore, $t_H \leq |\rho_H(x)|$ by minimality of $t_H$.
	This proves item~\ref{prop:en3}.
	
	Let $u \in H$. Let $t' = |\rho_u(x)|$.
	By induction, let us prove that all vertices $v$ such that there is a path of length $d$ from $u$ to $v$, we have $\rho_v(x) = \sigma^{i}( \rho_u(x))$ with $i=d \bmod t'$.
	It is true for $d=0$ because $\rho_u(x) = \sigma^{0}( \rho_u(x))$.
	Suppose this is true for $d$. 
	Let $v \in H$ with a path $(u,\dots,v',v)$ of length $d+1$ from $u$.
	Then, there exists a path $(u,\dots,v')$ of size $d$ between $u$ and $v'\in \inNeighbors_H(v)$.
	Therefore, $\rho_{v'}(x) = \sigma^{d \bmod t'}( \rho_u(x) )$.
    Now, by item~\ref{prop:en0}, $\rho_{v}(x) = \sigma(\rho_{v'}(x)) =  \sigma( \sigma^{d \bmod t'}( \rho_u(x) ) )$.
    Therefore, $|\rho_{v}(x)| = t'$ and $\rho_{v}(x) = \sigma^{ (d+1) \bmod t'}( \rho_u(x) )$.
    Now,  one can see that $t'=t_H$.
    Indeed, since the trace of all $v\in H$ are of same length $t'$, we have $f^{t'}_H(x) = x_H$ and therefore $t_H \leq t'$.
    Furthermore, by item~\ref{prop:en3}, $t' = |\rho_u(x)| \leq |\rho_H(x)| \leq t_H$.
    This proves item~\ref{prop:en4}.
	
	To prove item~\ref{prop:en5}, it is sufficient to fix $u \in H$ and to partition $H$ into $t_H$ parts $H_0, \dots, H_{t_H-1}$ where $H_i = \{ v \in H : \rho_v(x) = \sigma^i(\rho_u(x)) \}$ (the parts are disjoint and their union is $H$ by item~\ref{prop:en4}).
	If $v \in V_i$ and $v' \in \outNeighbors_H(v)$, then by item~\ref{prop:en0}, $\rho_{v'}(x) = \sigma(\rho_v(x)) = \sigma^{i+1}(\rho_u(x))$ and therefore $v' \in H_{i+1 \bmod t_H}$.
	
	To prove item~\ref{prop:en6}, it is sufficient to see that for all $v \in H_i$, we have $ x_v = (\rho_u(x))_i$.
	
	To prove item~\ref{prop:en7}, by item~\ref{prop:en6}, for all parts $H_i$ we have either $x_{H_i} = 0^{|H_i|}$ or $x_{H_i} = 1^{|H_i|}$.
	Now, if for all parts $H_i,$ we have $x_{H_i} = 0^{|H_i|}$ then $x_{H} = 0^{|H|}$, $f_H(x) = 0^{|H|} = x_H$ and $t=1$.
	The same reasoning can be used if for all parts $H_i$ we have $x_{H_i} = 1^{|H_i|}$ and this proves item~\ref{prop:en7}.
	
	To prove item~\ref{prop:en8}, it is sufficient to see that $k$ is the length of the periodic trace $\rho_V(x)$ which is a least common multiple of the length of the periodic traces of all $v \in V$.
	Since, $t_H$ is the length of the periodic trace $\rho_u(x)$ then $t_H$ divides $k$.
	
	To prove item~\ref{prop:en9}, it is sufficient to prove that $t_H$ divides the length of any cycle $c$ of $H$ (since $\cyclicity(H)$ is the $\lcm$ of all these lengths).
	Let $c$ be a cycle of $H$. Let $v \in c$. The cycle $c$ is a path of length $|c|$ from $v$ to $v$. Therefore, by item~\ref{prop:en4}, $\rho_v(x) = \sigma^i( \rho_v(x) )$ where $i = |c| \bmod t_H$.
	If $|c|$ does not divide $t_H$ then $\rho_v(x) = \sigma^i( \rho_v(x) )$ with $0<i<t_H$ with contradicts the minimality of $\rho_v(x)$.
	Therefore, $|c|$ divides $t_H$ and this proves item~\ref{prop:en9}.
	
	\qed

\end{pf}

The following result is straightforward from {\cite{AndOr08}}.
\begin{thm}\label{thm:PLCE_s_P}
	The PLCE problem (and therefore the $k$-PLCE problem) is in P when we restrict to conjunctive networks $f$ such that $D^f$ is strongly connected. 
\end{thm}
\begin{pf}
Consider the instance of the PLCE problem with $f$ a conjunctive network such that $D^f$ is strongly connected and an integer $k$.
Then by \refer{Lemma}{lem:plce} it is sufficient to compute the index of cyclicity $\ell$ of $D^f$ and then to check that $k$ divides $\ell$.
\end{pf}
In order to prove the NP-completeness of PLCE problem in the general case we    previously prove some results in Lemmas~\ref{lem:f^k(x)=x}~\ref{lem:local_period}~\ref{lem:xinPhiK} and Proposition~\ref{prop:primes} below.

Lemma~\ref{lem:f^k(x)=x} below states that it is polynomial to check if a given configuration is part of a limit cycle of a given length $k$.
\begin{lem} \label{lem:f^k(x)=x}
	For all conjunctive global functions $f$, for all integers $k$ and for all configurations $x \in \{0,1\}^n$, it is polynomial to check if $f^k(x) = x$.
\end{lem}
\begin{pf}
	Remember that by \refer{Lemma}{lem:effectivearcs}, the composition of conjunctive global functions is a conjunctive global function and that a conjunctive global function can be represented efficiently (for example, with its interaction digraph).
	First, even if $k$ is exponential in $n$, it is polynomial to compute $f^k$.
	Indeed, one can write $k$ in a sum of a logarithmic number of powers of 2: $k = 2^{k_1} + 2^{k_2} + \dots + 2^{k_m}$ and we have $f^k = f^k = f^{k_1} \circ f^{k_2} \circ \dots \circ f^{k_m}$.
	Furthermore, any function $f^{k_i}$ can be computed in a logarithmic number of compositions.
	For example,  $f^2 = f \circ f$, $f^4 = f^2 \circ f^2$, \textit{etc.}
	As a result, $f^k$ can be computed in polynomial time and it is then sufficient to check that $f^k(x) = x$ which is also polynomial (and even linear). \qed
\end{pf}

%

Lemma~\ref{lem:local_period} below show that if a configuration is in a limit cycle of length $k$, then $k$ is exactly the $\lcm$ of the periods of the limit not-trivial strongly connected components of $D^f$ (the period of a not-trivial strongly connected component $H$ being the smallest integer $t$ such that $(f^t(x))_H = x_H$ by Proposition~\ref{prop:ntscc}, item~\ref{prop:en3}).

\begin{lem} \label{lem:local_period}
	Let $H_1, \dots, H_m$ be the not trivial strongly connected components of $D^f$.
	Let $x \in \Phi(f)$ and for all $i \in [m],$ let $t_i>0$ be the smallest integer such that $x_{H_i} = (f^{t_i}(x) )_{H_i}$.
	Then $x \in \Phi_k(f)$ with $k = \lcm(t_1,\dots,t_m)$.
\end{lem}
\begin{pf}
    In Proposition~\ref{prop:ntscc}, item~\ref{prop:en3}, it is said that the periodic trace $\rho_{H_i}(x)$ is of length $t_i$.
    To prove that the periodic trace $\rho(x)$ is of length $k$, it is then sufficient to prove that the period $t_v = |\rho_v(x)|$ of each vertex $v$ out of a not trivial strongly connected component is a divisor of $k$.
    Consider an order of the vertices of $[n]$ not in a not trivial strongly connected component such that if there is a path from $u$ to $v$ then $u < v$. 
    Consider the smallest $v$ such that $t_v = |\rho_v(x)|$ does not divides $k$.
    So, $q$ such that $f^{q+k}(x) \neq f$ 
    \qed
\end{pf}

A consequence of Lemma~\ref{lem:local_period} is that the length of a limit cycle of a conjunctive network cannot be divided by an integer strictly greater than $n$.
\begin{cor} \label{cor:prime}
For all conjunctive global functions $f$, for all prime $p > n$ and for all $q$ multiple of $p$, $\Phi_q(f) = \emptyset$.
\end{cor}
\begin{pf}
    Let $x \in \Phi_k(f)$ for any integer $k$. 
    Let $H_1, \dots, H_m$ be the not trivial strongly connected components of $D^f$ and let $t_i$ be the smallest integer such that $x_{H_i} = (f^{t_i}(x) )_{H_i}$ for all $1 \leq i \leq m$.
    By Lemma~\ref{lem:local_period}, 
    $k = \lcm(t_1,\dots,t_m)$. 
    Furthermore, by item~\ref{prop:en9} of Proposition~\ref{prop:ntscc}, for all $1 \leq i \leq m$, $t_i \leq |H_i| \leq n$ so $k$ is not a multiple of any prime $p \geq n$.
    \qed
\end{pf}

Since for any update schedule $s$, $f^s$ is a conjunctive network if $f$ is a conjunctive network, then the result of Corollary~\ref{cor:prime} remains true for any update schedule.

Another consequence of Lemma~\ref{lem:local_period} is that the maximum length of a limit cycle of any conjunctive network of length $n$ corresponds to the Landau's function $\landau(n)$, that is the largest lcm of numbers $t_1,\dots,t_m$ whose sum is $n$ and is corresponds to the \href{https://oeis.org/A000793}{OEIS sequence A000793}.

\begin{cor} \label{cor:landau}
Let $n \in \mathbb{N}^*$, and let $k$ be the greatest integer such that there exists a conjunctive network $f:\{0,1\}^n \to \{0,1\}^n$ such that $\Phi_k(f) \neq \emptyset$. Then $k = \landau(n)$.
\end{cor}
\begin{pf}
    First, let us prove that $k \leq \landau(n)$.
    Let $x \in \Phi_k(f)$ for any integer $k$. 
    Let $H_1, \dots, H_m$ be the not trivial strongly connected components of $D^f$ and let $t_i$ be the smallest integer such that $x_{H_i} = (f^{t_i}(x) )_{H_i}$ for all $1 \leq i \leq m$.
    By Lemma~\ref{lem:local_period}, 
    $k = \lcm(t_1,\dots,t_m)$. 
    Let $t = \sum\limits_{1 \leq i \leq m} t_i$.
    By item~\ref{prop:en9} of Proposition~\ref{prop:ntscc}, for all $1 \leq i \leq m$, $t_i \leq |H_i|$ so $t \leq n$.
    Since, the Landau's function is an increasing function then $k \leq \landau(t) \leq \landau(n)$.
    
    Now, let us prove that $k \geq \landau(n)$.
    We take $t_1,\dots,t_m$ a partition of $n$ such that $\lcm(t_1,\dots,t_m)=\landau(n)$.
    Now consider $f$ a conjunctive network such that $D^f$ is composed of $m$ disjoint cycles $c_1, \dots, c_m$ such that $|c_i|=t_i$ for all $1 \leq i \leq m$.
    We take $x \in \{0,1\}^n$ such that $x$ has exactly one $1$ in each cycle $c_i$.
    So the period trace $\rho_{H_i}$ is of length $t_i$.
    As a result, $x \in \phi_k(x)$ with $k = \lcm(t_1,\dots,t_m)=\landau(n)$.
    
    \qed
    
\end{pf}

More generally, the lengths of all the limit cycle of all conjunctive networks of length $n$ are exactly all the $\lcm$ of numbers $t_1,\dots,t_m$ whose sum is $n$.

\begin{lem} \label{lem:xinPhiK}
	For all conjunctive global functions $f$, for all configuration $x \in \{0,1\}^n$, for all integer $k$, it is polynomial to check if $x \in \Phi_k(f)$.
\end{lem}
\begin{pf}
	By \refer{Lemma}{lem:f^k(x)=x}, it is polynomial to check that $f^k(x) = x$.
	It is not sufficient to conclude that $x \in \Phi_k(f)$ because it could exist an integer $1 \leq q < k$ such that $f^q(x) = x$, but it proves that $x \in \Phi(f)$.
	Next, one can decompose $D^f$ into no trivial strongly connected components $H_1, \dots, H_m$ and compute $t_1, \dots t_m$, the smallest strictly positive integers such that $(f^{t_i}(x) )_{H_i} = x_{H_i}$.
	Any integer $t_i$ can be computed in polynomial time because by Proposition~\ref{prop:ntscc}, item~\ref{prop:en8}, $t_i$ is smaller than $n$.
	By \refer{Lemma}{lem:local_period}, $x \in \Phi_{\lcm(t_1,\dots,t_m)}(f)$ and it is then sufficient to check that $k = \lcm(t_1,\dots,t_m)$.
	\qed
\end{pf}

Proposition~\ref{prop:primes} below show that if there is a path between two non-trivial strongly connected component $H_1$ and $H_2$ then in all limit cycle, either one of the component is stable or the two component's period are not co-prime.
\begin{prop} \label{prop:primes}
Let $f$ be a conjunctive global function with at least two strongly connected components $H_1$ and $H_2$ with a path between $H_1$ and $H_2$.
Let $x \in \Phi(f)$.
Then either $t_{H_1} = 1$ or $t_{H_2}=1$ or $\gcd(t_{H_1},t_{H_2})\neq1$ with $t_{H_1}$ and $t_{H_2}$ the smallest integers such that $f^{t_{H_1}}_{H_1}(x) = x_{H_1}$ and $f^{t_{H_2}}_{H_2}(x) = x_{H_2}$.
\end{prop}
\begin{pf}
Suppose that $t_{H_1} \neq 1$ and $\gcd(t_{H_1},t_{H_2}) = 1$.
Let $u \in H_1$ and $v \in H_2$.
Since $t_{H_1} \neq 1$, by item~\ref{prop:en7} of proposition~\ref{prop:ntscc}, there exists $u \in H_1$ such that $x_u = 0$ and by item~\ref{prop:en4}, for all $k \geq 0$, $f^{k t_{H_1}}_u(x)=0$.
Now, consider the path $u_0=u,u_1,\dots,u_{\ell}=v$ between $u$ and $v$ with $v\in H_2$.

By induction, one can show that for all $k\geq0$, 
$f^{k t_1 + i}_{u_i}(x) = 0$.
Indeed, $f^{k t_1+0}_{u_0}(x) = f^{k t_1}_{u}(x) = 0$ and for all $i>0$, $f^{k t_1 +i}_{u_i}(x) =  f^{k t_1 +(i-1)}_{u_{i-1}}(x) \wedge \dots = 0 \wedge \dots = 0$.
As a result, for all $k \geq 0,$
$f^{k t_1+\ell}_v(x) = 0$.

Now, for the sake of contradiction, suppose that $t_{H_2}\neq1$.
As a result, there exists $q$ such that for all $k'\geq0$,
$f^{k' t_{H_2} + q}_v(x) = 1$.
However, since $t_1$ and $t_2$ are co-primes then there exists $k$ and $k'$ such that $k t_{H_1}+\ell=k' t_{H_2}+q$ and therefore, for this couple $(k,k)'$, $f^{k t_{H_1}+\ell}_v(x) =0$ and $f^{k t_{H_1}+\ell}_v(x) =1$.
This is a contradiction. 
Therefore, $t_{H_2}=1$.
\qed
\end{pf}

Theorem~\ref{thm:SLCE-NP} below shows that the general problem of knowing if a conjunctive network has a limit cycle of length $k$, with $k$ an entry of the problem given in binary is NP-complete.
\begin{thm} \label{thm:SLCE-NP}
	The PLCE problem is NP-complete in general. 
\end{thm}
\begin{pf}
	
	In this proof, the prime numbers are denoted by $p_1, p_2, \dots$. Consider a conjunctive global function $f$ and an integer $k$.
	To prove that $\Phi_k(f) \neq \emptyset$, is is sufficient to exhibit a configuration $x \in \Phi_k(f)$.
	By \refer{Lemma}{lem:xinPhiK}, it is then polynomial to check that $x$ really belongs to $\Phi_k(f)$.
	Therefore, the PLCE problem is in NP.
	
	Now, let us prove that the PLCE problem in NP-hard.
	For this purpose we will reduce from 3-SAT. 
	Consider a 3-CNF formula $\psi$ composed of $n$ variables $\lambda = \{ \lambda_1, \dots, \lambda_n\}$ and $m$ clauses $\mu = \{ \mu_1, \dots, \mu_m \}$.
	Consider the following conjunctive global functions $f$ described by its digraph $H$.
	First, each variable $\lambda_i$ is represented by two isolated cycles of length $p_i$.
	We will refer to the first one as $\Lambda_{i,\top}$ and the second one as $\Lambda_{i,\bot}$.
	For example, $\lambda_2$ is represented by two cycles of length $p_2 = 3$.
	Second, each clause $\mu_j$ is represented by three cycles of length $p_{n+j}$ referred as $M_{j,1}$, $M_{j,2}$ and $M_{j,3}$ (each one corresponding to a literal of the clause).
	Now, for each literal $\mu_{j,\ell} = (\lambda_i, \rho)$ with $j \in [m], \ell \in [3], i \in [n]$ and $\rho \in \{ \top, \bot \}$,
	there are two cases.
	If $\rho = \top$, then there is an arc from the component $\Lambda_{i,\top}$ to the component $M_{j,\ell}$.
	Otherwise,  then there is an arc from the component $\Lambda_{i,\bot}$ to the component $M_{j,\ell}$.
	Finally, let $k = p_1 p_2 \dots p_{n+m}$.
	
	This complete the description of this reduction. 
	Note that $p_{n+m} \leq (n+m) (\ln (n+m) + \ln \ln (n+m)) \leq (n+m)^2$ for $(n+m)$ big enough~\cite{robin1983estimation}.
	Furthermore, there are two cycles of size of length $\leq p_{n+m}$ for each variable and at most $3$ cycles of length $\leq p_{n+m}$.
	So $D^f$ is of polynomial size $\leq 3(n+m)^3$.
	
	We claim that $\Phi_k(f) \neq \emptyset$ if and only if $\psi$ has a solution.

	First, suppose that there exists a valid valuation $v: \lambda \to \{\top, \bot\}$ of $\psi$.
	Then we consider the following configuration $x$.
	First, for all $i \in [n]$, there are two cases.
	\begin{itemize}
	\item if $v( \lambda_i) = \top$, then $x_{\Lambda_{i,\bot}} = 1 (0)^{p_i-1}$ and $x_{\Lambda_{i,\top}} = (0)^{p_i}$, and
	\item if $v( \lambda_i) = \bot$, then $x_{\Lambda_{i,\bot}} = (0)^{p_i}$ and $x_{\Lambda_{i,\top}} = 1 (0)^{p_i-1}$.
	\end{itemize}

	Similarly, for any literal $\mu_{j,\ell} = (\lambda_i, \rho)$, 
	if $v( \mu_{j,\ell}) = \top$ (\textit{i.e.} if $v(\lambda_i) = \rho$), then $x_{M_{j,\ell}} = 1 (0)^{p_{n+j}-1}$ and $x_{M_{j,\ell}} = (1)^{p_{n-j}}$ otherwise.
	We will now prove that if $v$ is a valid valuation then $x \in \Phi_k(f)$.
	
	For all, $H$ a strongly connected component of $D$ (\textit{i.e.} $H = \Lambda_{i,\rho}$ or $H = M_{j,\ell}$), 
	let $t_H$ be the smallest strictly positive integer such that $f^{t_H}(x) = x_H$ (we prove that such an integer exists all the time next).
	Let $T$ be the set of all these integers.
	Let $i \in [n]$ and $\rho = v(\lambda_i)$.
	Note that the cycles $\Lambda_{i,\top}$ and $\Lambda_{i, \bot}$ are isolated: 
	the only in-neighbors of the vertices are their predecessors in the cycles.
	As a consequence we have 
	\begin{align*}
	x_{\Lambda_{i,p} } &= 1 (0)^{p_i-1}  \\
	(f(x))_{\Lambda_{i,p} } &= 0 1 (0)^{p_i-2} \neq x_{\Lambda_{i,p} } \\
	&\dots\\
	(f^{p_i-1}(x))_{\Lambda_{i,p} } &= (0)^{p_i-1} 1 \neq x_{\Lambda_{i,p} }  \\
	(f^{p_i}(x))_{\Lambda_{i,p} } &= 1 (0)^{p_i-1} = x_{\Lambda_{i,p} } .
	\end{align*}
	and
	\begin{align*}
	(f(x))_{\Lambda_{i,\neg p} } &= (0)^{p_i} &= x_{\Lambda_{i,\neg p} }.
	\end{align*}
	We can see that if $v(\rho) = \top$ then $t_{\Lambda_{i,\top}} = 1$ and $t_{\Lambda_{i,\bot}} = p_i$.
	Otherwise, $t_{\Lambda_{i,\top}} = p_i$ and $t_{\Lambda_{i,\bot}} = 1$.
	In both cases, we have $\lcm(t_{\Lambda_{i,\top}},t_{\Lambda_{i,\bot}}) = p_{i}$.
	
	The same way, for all $\mu_{j,\ell} = (\mu_i,\rho) \in \mu$, their only in-neighbors are their predecessor in the cycle $M_{j,\ell}$  and, for one of them, a vertex of $\Lambda_{i,\rho}$.
	They are two cases: 
	If $v(\mu_{j,\ell})=\top$ (\textit{i.e.} $v(\lambda_i) = \rho$) then 
	\[
	x_{\Lambda_{i,\rho}} = (f(x))_{\Lambda_{i,\rho}} = (f^2(x))_{\Lambda_{i,\rho}} = \dots = (0)^{p_i}.
	\]
	Furthermore,
	\begin{align*}
		x_{M_{j,\ell}} &= 1 (0)^{p_{j+n}-1} \\
		(f(x))_{M_{j,\ell}} &= 0 1 (0)^{p_{j+n}-2} \neq x_{M_{j,\ell}} \\
		& \dots \\
		(f^{p_{j-n}}(x))_{M_{j,\ell}} &= 1 (0)^{p_{j+n}-1} = x_{M_{j,\ell}}
	\end{align*}
	
	Therefore, we have $t_{M_{j,\ell}} = p_{j+n}$.
	Otherwise, if $v(\mu_{j,\ell})=\bot$, then $x_{M_{j,\ell}} = (1)^{p_{j+n}}$ and $(f(x))_{M_{j,\ell}} = (1)^{p_{j+n}}$ and we have $t_{M_{j,\ell}} = 1$.
	
	Note that since $v$ is a valid valuation of $\psi$, we have $v(\mu_{j,1}) = \top$ or $v(\mu_{j,2}) = \top$ or $v(\mu_{j,3}) = \top$.
	This signifies that $\lcm(t_{M_{j,1}}, t_{M_{j,2}}, t_{M_{j,3}}) = p_{j+n}$.
	As a result,
	\[
	\lcm(T) = {\displaystyle \prod_{i=1}^{n} \lcm(t_{\Lambda_{i,\top}},t_{\Lambda_{i,\bot}}) . \prod_{j=1}^{m} \lcm(t_{M_{j,1}}, t_{M_{j,2}}, t_{M_{j,3}})} =  {\displaystyle \prod_{i=1}^{n} p_{i} . \prod_{j=1}^{m} p_{j+n}} =  {\displaystyle \prod_{i=1}^{n+m} p_{i} } = k.
	\]
	 Hence, $f^k(x) = x$, $x \in \Phi(f)$.
	 Now, by \refer{Lemma}{lem:local_period}, we have $x \in \Phi_q(f)$ with $q = \lcm(T) = k$.
	 As a result, $x \in \Phi_k(f)$.
	 
	 In the other hands, suppose that there exists $x \in \Phi_k(f)$.
	 Let us prove that there exists a valid valuation $v:\lambda \to \{\top,\bot\}$.
	 First, \refer{Lemma}{lem:local_period}, there exists a set of integers $T = \{ t_H: H \text{ a strongly connected component of } D^f \} $ such that for all $t_H \in T,$ $t_H$ is the smallest strictly positive integer such that $(f^{t_H}(x))_H = x_H$.
	 By item~\ref{prop:en6} of proposition~\ref{prop:ntscc}, for all $i\in[n]$ and $\rho \in \{ \top, \bot \},$ $t_{\Lambda_{i,\rho}}$ must divides $\cyclicity(\Lambda_{i,\rho}) = p_i$. Since $p_i$ is prime then $t_{\Lambda_{i,\rho}} = 1$ or $t_{\Lambda_{i,\rho}} = p_i$.
	 The same way,  for all $j\in[m]$ and $\ell \in [3]$, $t_{m_{j,\ell}} = 1$ or $t_{m_{j,\ell}} = p_{j+n}$.
	
	Second, for all $i \in [n]$ and for all $\rho \in \{ \top, \bot \},$ we have $t_{\Lambda_{i,\top}} = p_i$ or $t_{\Lambda_{i,\bot}} = p_i$ because otherwise $p_i$ is not a factor of $k = {\displaystyle \prod_{i=1}^{n+m} p_{i} }$ which is a contradiction.
	The same way, for all $j\in[m]$, we have $t_{M_{j,1}} = p_{j+n}$ or $t_{M_{j,2}} = p_{j+n}$ or $t_{M_{j,3}} = p_{j+n}$.
	
	We define $v: \lambda \to \{ \top, \bot \}$ such that for all $i \in [n],$ $v(\lambda_i) = \top$ if	$t_{\Lambda_{i,\bot}} = p_i$ and $v(\lambda_i) = \bot$ otherwise (note that in this second case we have $t_{\Lambda_{i,\top}} = p_i$).
	Let us prove that $v$ is a valid valuation of $\psi$.
	
	Let $\mu_{j,\ell} = (\lambda_i,\rho) \in$ be a literal such that $v(\mu_{j,\ell}) = \bot$ (\textit{i.e.} $v(\lambda_i) \neq \rho$).
	Then, we have a path from $\Lambda_{i,\rho}$ to $M_{j,\ell}$ (by definition of $D^f$).
	By proposition~\ref{prop:primes}, $t_{\Lambda_{i,\rho}} =1$ or $t_{M_{j,\ell}} =1$ or $\lcm(t_{\Lambda_{i,\rho}},t_{M_{j,\ell}}) \neq 1$.
	We know that $t_{M_{j,\ell}} = 1$ or $t_{M_{j,\ell}} = p_{j+n}$ and $t_{\Lambda_{i,\rho}} = p_i$ (by definition of $v$).
	The primes $p_i$ and $p_{j+n}$ are different and therefore $\lcm(t_{\Lambda_{i,\rho}},t_{M_{j,\ell}}) \neq 1$.
	As a result, $t_{\Lambda_{i,\rho}}$ must be equal to $1$.
	This signifies that for all literal $\mu_{j,\ell}$, $v(\mu_{j,\ell}) = \bot$ implies $t_{\Lambda_{i,\rho}} = 1$.
	
	Therefore, if $v$ is not a valid valuation, then there exists $j \in [m]$ such that $v(\mu_{j,1}) = v(\mu_{j,2}) = v(\mu_{j,3}) = \bot$ and therefore $t_{M_{j,1}} = t_{M_{j,2}} = t_{M_{j,3}} = 1$.
	This is a contradiction because we already proved that  $t_{M_{j,1}} = p_{j+n}$ or $t_{M_{j,2}} = p_{j+n}$ or $t_{M_{j,3}} = p_{j+n}$.
	As a result, $v$ is a valid valuation of $\psi$.
	
	This concludes the proof that PLCE is NP-hard and therefore NP-complete.
	
	\qed
\end{pf}

A question still open is the complexity of the $k$-PLCE problems when the interaction digraph is not strongly connected.

We know by Lemma~\ref{lem:local_period} and item~\ref{prop:en5} of Proposition~\ref{prop:ntscc}, that a necessary condition for $k$-PLCE($f$) to have a solution is that $k$ equals $\lcm(t_1,\dots,t_m)$ for $t_1,\dots,t_m$ dividing respectively $c(H_1), \dots, c(H_m)$ with $H_1, \dots, H_m$ the strongly connected component of $D^f$.

For some values of $k$, this condition is sufficient and that makes the problem in P as shown in Proposition~\ref{prop:power_of_prime} below.
\begin{prop}\label{prop:power_of_prime}
    If $k$ is a power of a prime, then $k$-PLCE($f$) has a solution if and only if $D^f$ has a strongly connected component $H$ such that $k$ divides $\cyclicity(H)$.
\end{prop}
\begin{pf}
    The left to right direction is a direct consequence of the Lemma~\ref{lem:local_period} and item~\ref{prop:en5} of Proposition~\ref{prop:ntscc}.
For the right to left direction, using the strongly connected component $H$, we can create a configuration $x \in \Phi_k(f)$ the following way. 
First, we know that $H$ is $k$-partite for a partition $H^1,\dots,H^k$.  
We take $x_{H^1} = 1^{|H^1|}$ and $x_{H \setminus H^1} = 0^{|H \setminus H^1|}$ and we have $t_H = k$.
For all the other strongly connected components $H'$ we take $x_{H'} = 0^{|H'|}$ and we have $t_{H'}=1$.
For the vertices $v$ out of any strongly connected component, if there is a path from a strongly connected component other than $H$ to $v$ then $x_v=0$.
Else $x_v = 0$ iff and only if there is path between $H^1$ and $v$ and all path between $H^1$ and $v$ have lengths multiple of $k$.
One can check that $x \in \Phi_k(f)$.
\qed
\end{pf}

However, there are more complicated cases.
For example, by Proposition~\ref{prop:primes}, if $D^f$ is composed of 2 cycles of length $2$ and $3$ connected by an arc then $f$ has no limit cycle of size $6$ because $2$ and $3$ are co-primes. However, if $D^f$ is composed of 2 cycles of length $6$ and $10$ connected by an arc, then $f$ has a limit cycle of length $30$. The exact characterization of when $k$-PLCE(f) has a solution is unknown and, therefore it is an open problem to know for which values of $k$, $k$-PLCE is in P.

\section{Complexity of the BLCE, SLCE, $k$-BLCE and $k$-SLCE problems} \label{section:BLCE}

In this section, we study the problem of the existence of limit cycle of a given length for a block-sequential or sequential update schedule.  

\begin{lem}[\cite{Math12}] \label{lem:parallel}
    Let $f$ be a conjunctive network such that $D^f$ is strongly connected and let $s$ be a block-sequential update schedule.
    Then, $\mathcal{P}(D^f_s)$ has a unique strongly connected component.
\end{lem}

\begin{prop}\label{lem:cyclesGlab}
	Let $D_{\lab}$ be an update digraph. Then, every circuit in $D_{\lab}$ produces a circuit in $\mathcal{P}(D_{\lab})$ with length the number of the $\mi$-labeled arcs of it. Conversely, every circuit in $\mathcal{P}(D_{\lab})$ comes from a circuit in $D_{\lab}$ with a number of $\mi$-labeled arcs equal to the length of the cycle.
\end{prop}

\begin{pf}
Let us consider $D=(V,A)$ and $\mathcal{P}(D_{\lab})=(V,A')$.

Let $c = (c_1, \dots, c_m)$ a circuit in $D_{\lab}$ with $\ell$ $\mi$-labeled arcs.
Let us prove that there is a circuit of length $\ell$ in $D$.
Let $c_{i_1}, \dots, c_{i_\ell}$ be the $\ell$ vertices of $c$ such that $(c_{i_j-1},c_{i_j}$) is a $\mi$-labeled arc for all $1 \leq j \leq \ell$ with $i_1 \leq i_2 \leq \dots \leq i_\ell$.
For all $1 \leq j \leq m$, such that $i_j+1 < i_{j+1} $,
we have $\lab (c_{i_j},c_{i_j+1}) = \dots = (c_{i_{j+1}-2},c_{i_{j+1}-1}) = \me$ and 
$(c_{i_{j+1}-1},c_{i_{j+1}}) = \mi$.
Hence, by definition of $\mathcal{P}(D_{\lab})$, $(c_{i_j},c_{i_{j+1}}) \in \mathcal{P}(D_{\lab})$ and $(c_{i_1}, \dots, c_{i_\ell})$ is a circuit in $\mathcal{P}(D_{\lab})$ which is also a cycle if $c$ is a cycle.

In the other hand, take $(c_1,\dots,c_\ell)$ a circuit in $\mathcal{P}(D_{\lab})$.
Then, by definition of $\mathcal{P}(D_{\lab})$, for all $1 \leq j \leq \ell$, $\lab(c_j,c_{j+1})=\mi$ or there is a path $(c_j = v_1,v_2, \dots, c_{j+1} = v_r)$ in $D_{\lab}$ such that $\lab (v_1,v_2) = \dots = (v_{r-2},v_{r-1}) = \me$ and 
$(v_{r-1},v_r) = \mi$.
As a result, there is a circuit in $D_{\lab}$ with exactly $\ell$ $\mi$-labeled arcs.
	\qed
\end{pf}

\begin{prop} \label{prop:klab}
	Let $f$ be a global conjunctive function and $s$ be a block-sequential update schedule network with strongly connected interaction digraph.
	Then, $\Phi_k(f^s) \neq \emptyset$ if and only if $D_s^f$ is $k$-labeled.
\end{prop}
\begin{pf}

    First, by Proposition~\ref{lem:parallel},  $D^{f^s} = \mathcal{P}(D^f_s)$ has a unique strongly connected component.
    
    Suppose that $D_s^f$ is $k$-labeled. 
    Then 
    every circuit in $D_s^f$ has a multiple of $k$ $\mi$-labeled arcs.
    Therefore, every circuit in $\mathcal{P}(D^f_s)$ is of length multiple of $k$. Indeed, if there is a circuit of length no multiple of $k$ in $\mathcal{P}(D^f_s)$ then there is a circuit in $D_s^f$ with a number of $\mi$-labeled arcs no multiple of $k$ which is a contradiction by Proposition~\ref{lem:cyclesGlab}.
    Since any circuit in $\mathcal{P}(D^f_s)$ has a length multiple of $k$ then $k$ divides the index of cyclicity of the strongly connected component.
    As a result, by Lemma~\ref{lem:plce}, $\Phi_k(f^s) \neq \emptyset$.
    
    The other direction is symmetrical. Suppose that $\Phi_k(f^s) \neq \emptyset$.
    Then, by Proposition~\ref{lem:plce} the index of cyclicity of the strongly connected component of $D^{f^s}$ is a multiple of $k$.
    Therefore, every circuit in $D^{f^s}$ is of length multiple of $k$ and by Proposition~\ref{lem:cyclesGlab}, every circuit in $D^f_s$ has a number of $\mi$-labeled multiple of $k$.
    Then $D^f_s$ is $k$-labeled.
    \qed
   
\end{pf}


\begin{prop} \label{prop:solution}
	Suppose that $f$ is a conjunctive global function such that $D^f$ is strongly connected.
	Then, $k$-BLCE($f$) has a solution if and only if there exists a $k$-labeling $\lab$ of $D^f$ such that $D^f_{\lab}$ is an update digraph.
	Furthermore, $k$-SLCE($f$) has a solution if and only if there exists a $k$-labeling of $D^f$ $\lab$ such that $D^f_{\lab}$ is a sequential update digraph.
\end{prop}
\begin{pf}
    $k$-BLCE($f$) (\textit{resp.} $k$-SLCE($f$)) has a solution $\Leftrightarrow$ there exists a block sequential update schedule $s$ such that $\Phi_k(f^s) \neq \emptyset$ $\Leftrightarrow$ $D^f_s$ is an update $k$-labeled digraph (sequential update $k$-labeled digraph if $s$ is sequential)
    
    \qed
\end{pf}

To prove Theorem~\ref{thm:impl} below which states that if $k$-BLCE($f$) has a solution then so do $k$-SLCE($f$), we first need to introduce the concept of \emph{reversed} digraph which permit a characterization of the (sequential) update digraph. 

Given a labeled digraph $D_{\lab}=(D=(V,A),\lab)$, we define \mbox{$D_{\lab}^{R}=(D^R=(V,A^R),\lab^R)$}, the \emph{reverse digraph}, as follows:

\begin{itemize}
\item $(u,v)\in A^R\ssi ((u,v)\in A\land \lab(u,v)=\mi)\lor((v,u)\in A\land \lab(v,u)=\me)$.
\item $\lab^R(u,v)=\me$ if $\lab(v,u)=\me$ and $\lab^R(u,v)=\mi$ otherwise.
\end{itemize}

Basically, each $\me$-arc $(u,v)$ in $D_{\lab}$ gives a $\me$-arc $(v,u)$ in $D_{\lab}^{R}$ and each $\mi$-arc $(u,v)$ in $D_{\lab}$ gives a $\mi$-arc $(u,v)$ in $D^R_{\lab}$ if $D^R$ not already have a $\me$-arc $(u,v)$.
See an example of a interaction
digraph in \refer{\figurename}{Fig:RevDig}).

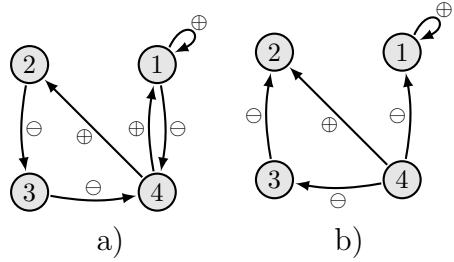
\begin{figure}[h]
\centering
\input{FigEjRevD}
\caption[Example of a reverse digraph.]{In a) a partial labeled digraph and in b) its reverse digraph.}
\label{Fig:RevDig}
\end{figure}

A cycle in $D_{\lab}^{R}$ is called \emph{forbidden} if it contains at least one $\me$-arc. 

It was shown in \cite{Math11} that there is a characterization of the (sequential) update digraph by the reverse digraph.

\begin{lem} [\cite{Math11}] \label{lem:car}
$D_{\lab}$ is an update digraph if and only if $D_{\lab}^{R}$ does not contain any forbidden cycle. Furthermore, $D_{\lab}$ is a sequential update digraph if and only if $D_{\lab}^{R}$ is acyclic.
Besides, these properties can be tested in polynomial time.
\end{lem}

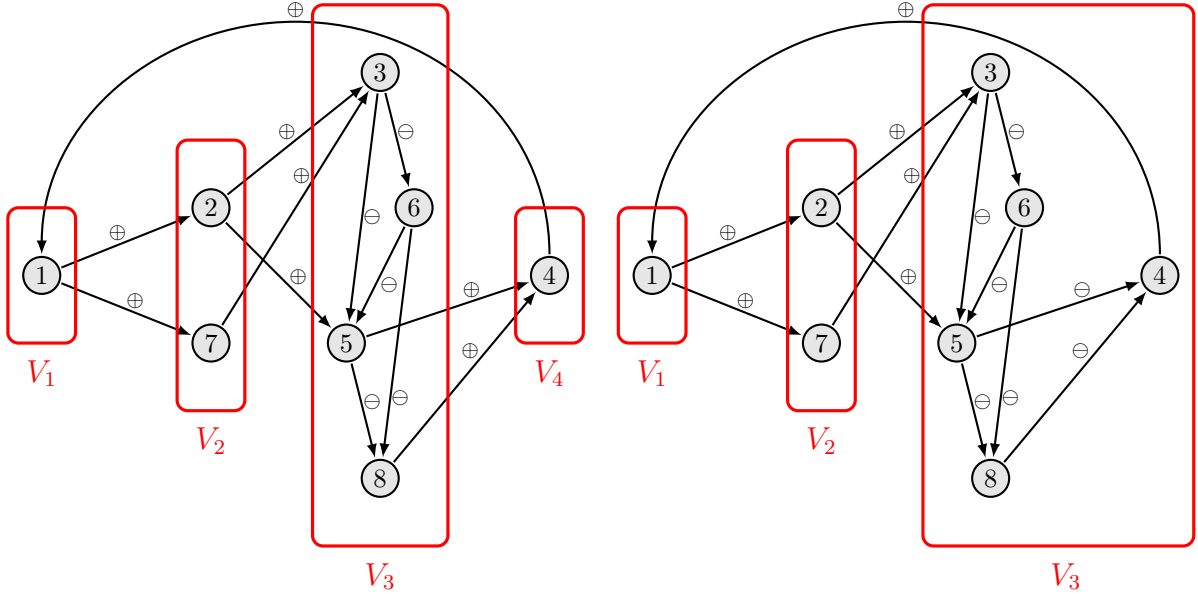
\begin{figure}[h]
\centering
\input{PGlabyGlabOrdered}
\caption{If $D_s$ is $k$-labeled update schedule then exists $D_{s'}$ $(k-1)$-labeled sequential update schedule}\label{fig:part2}
\end{figure}

\begin{thm} \label{thm:impl}
	For any conjunctive global function $f$ such that $D^f$ is strongly connected and for any $k \geq 3$,
	$k$-BLCE($f$) $ \implies$ $(k-1)$-SLCE($f$) $\implies$ $(k-1)$-BLCE($f$).
\end{thm}
\begin{pf}
	Let $k \geq 3$, and consider a conjunctive global function $f$ and $D^f = ([n],A)$.
	Suppose that $k$-BLCE($f$) (resp. $k$-SLCE($f$)) has a solution.
	Then there exists a labeling $\lab$ and a partition 
	$V_0, \dots, V_{k-1}$ of $[n]$ such that for every arc $(i,j) \in A$, if $i \in V_p$, we have $j \in V_{(p+1) \bmod k}$ if $(i,j)$ is a positive arc and $j \in V_{p}$ otherwise (and without cycle in $V_\ell$ for all $0 \leq \ell \leq k-1$).
	Therefore, a $(k-1)$-labeling $\lab'$ can be obtained from $\lab$ by switching all arc $(i,j)$ with $i \in V_{k-2}$ and $j \in V_{k-1}$ ($V_{k-2} \cup V_{k-1}$ is a part in the new partition) see Figure~\ref{fig:part2}. 
	Furthermore, the arcs from $V_{k-2}$ to $V_{k-1}$ is a feedback arc set of $D^f$, and all arcs from $V_{k-2}$ to $V_{k-1}$ are $\me$ arcs in $D^f_{\lab'}$. 
	Hence, $D^f_{\lab'}$ has no cycle with only $\mi$ arcs, and therefore it corresponds to a sequential update schedule by Lemma~\ref{lem:car}. 
	Therefore, $(k-1)$-SLCE($f$) has a solution.
	\qed
\end{pf}

\begin{figure}[h] \label{figure:BLCEnoSLCE}
	\begin{center} 
		\input{fig/BLCEnoSLCE}
	\end{center}
	\caption{Example of $D^f$ such that 2-BLCE(f) has a solution, but not 2-SLCE(f) and not 2-PLCE($f$).}
	\end{figure}
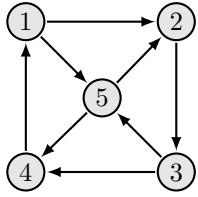

Figure~\ref{figure:BLCEnoSLCE} shows an example of $D^f$ such that 2-BLCE($f$) has a solution, but not 2-SLCE($f$) and not 2-PLCE($f$).

\begin{thm} \label{thm:NP_c}
	The BLCE and SLCE problems are NP-complete and for all $k \geq 2$, the problems $k$-BLCE and $k$-SLCE are NP-complete.
\end{thm}

Theorem~\ref{thm:NP_c} is a direct consequence of Lemmas~\ref{lem:NP} and \ref{lem:k-BLCE_hard} below.

\begin{lem}~\label{lem:NP}
	The BLCE, $k$-BLCE, SLCE and $k$-SLCE problems are in NP.
\end{lem}
\begin{pf}
	
	A possible certificate is to give a block-sequential (\textit{resp.} sequential) update schedule $s$ and a configuration $x \in \Phi_k(f^s)$.
	Reminder that $f^s$ is a conjunctive global function. 
	By \refer{Lemma}{lem:xinPhiK} it is polynomial to check that $x \in \Phi_k(f^s)$.
	\qed

\end{pf}

\begin{lem}[{\cite{macauley2009cycle}}] \label{lem:perm} 
	Let $f$ be a global function and $s= (B^1,\dots,B^{p-1},B^{p})$ be a block-sequential update schedule, 
	then for any $k \geq 1$, $|\Phi_k(f^s)| = |\Phi_k(f^{(B^p,B^1,\dots,B^{p-1})})|$.
\end{lem}

\begin{cor}
	If for one instance $f$, the BLCE (\textit{resp.} SLCE) problem has a solution with a block-sequential (\textit{resp.} sequential) update schedule $s$, then, for each block $X$ of $s$, there exist another solution $s'$ which update $X$ first.
\end{cor}
\begin{pf}
	Direct from Lemma~\ref{lem:perm}.
	\qed
\end{pf}

To prove that these problems are NP-hard, we do a reduction from 3-SAT.
We will first consider $k=2$ and then generalize.

\begin{lem} \label{lem:NP_hard_2}
	The $2$-BLCE and $2$-SLCE problems are NP-hard.
\end{lem}
\begin{pf}
	Consider a 3-CNF formula $\psi$ composed of $n$ variables $\lambda = \{ \lambda_1, \dots, \lambda_n \}$ and $m$ clauses $\mu = \{ \mu_1, \dots, \mu_m \}$.
	Construct the following digraph $ D = (V,A)$ whose answer to the $k$-BLCE and $k$-SLCE problems are positives if and only if the formula $\psi$ can be satisfied.
	
	First, each variable $\lambda_i \in \lambda$ is represented in $D$ by four vertices $x_i,t_i,\bar{x_i},f_i$ and five arcs $(x_i,t_i),(t_i,\bar{x_i}),(\bar{x_i},f_i), (f_i,x_i)$ and $(t_i,f_i)$ as shown in \refer{Figure}{figure:variable_D_psi}.
	Let $X = \{ x_i: i \in [n] \} \cup \{ \bar{x_i}: i \in [n] \}$.
	
	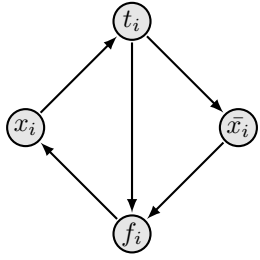
\begin{figure}[h] \label{figure:variable_D_psi}
	\begin{center} 
		\input{fig/variable}
	\end{center}
	\caption{Representation of a variable of a 3-CNF $\psi$ in the digraph $D^\psi$.}
	\end{figure}	
	Second, each clause $\mu_j \in \mu$ is represented by a cycle of seven vertices $c_{j,1}, \ell_{j,1}, c_{j,2}, \ell_{j,2}, c_{j,3}, \ell_{j,3}$ and $c_{j,4}$ (in this order) as represented in \refer{Figure}{figure:clause_D_psi}.
	Each node $\ell_{j,p}$ corresponds to a literal $\mu_{j,p} = ( \lambda_i, \rho)$.
	If $\rho = \top$, then we add the two arcs $(x_i,\ell_{j,p})$ and $(\ell_{j,p},\bar{x_i})$.
	Otherwise, we add two arcs $(\bar{x_i},\ell_{j,p})$ and $(\ell_{j,p},x_i)$.
	Let $C = \{ c_{j,p}: j \in [m] \text{ and } p \in [4] \}$ and 
	$L = \{ \ell_{j,p}: j \in [m] \text{ and } p \in [3] \}$.

	\begin{figure}[h] \label{figure:clause_D_psi}
	\begin{center} 
		\input{fig/clause}
	\end{center}
	\caption{Representation of a clause of a 3-CNF $\psi$ in the digraph $D^\psi$.}
\end{figure}
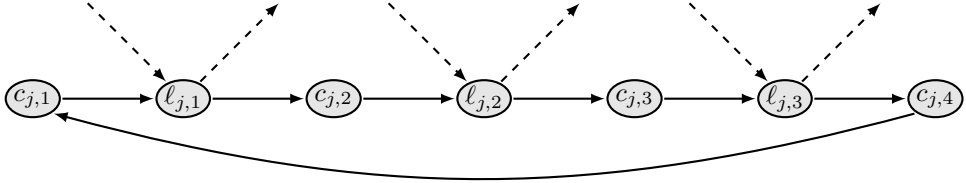	
	
	Finally, we add two vertices $a_1,a_k$, an arc $(a_1,a_k)$ and, for each vertex $v \in X \cup C$, we add two arcs $(v,a_1)$ and $(a_k,v)$. An example of digraph $D^\psi$ is represented \refer{Figure}{figure:reduction_D_psi}.

	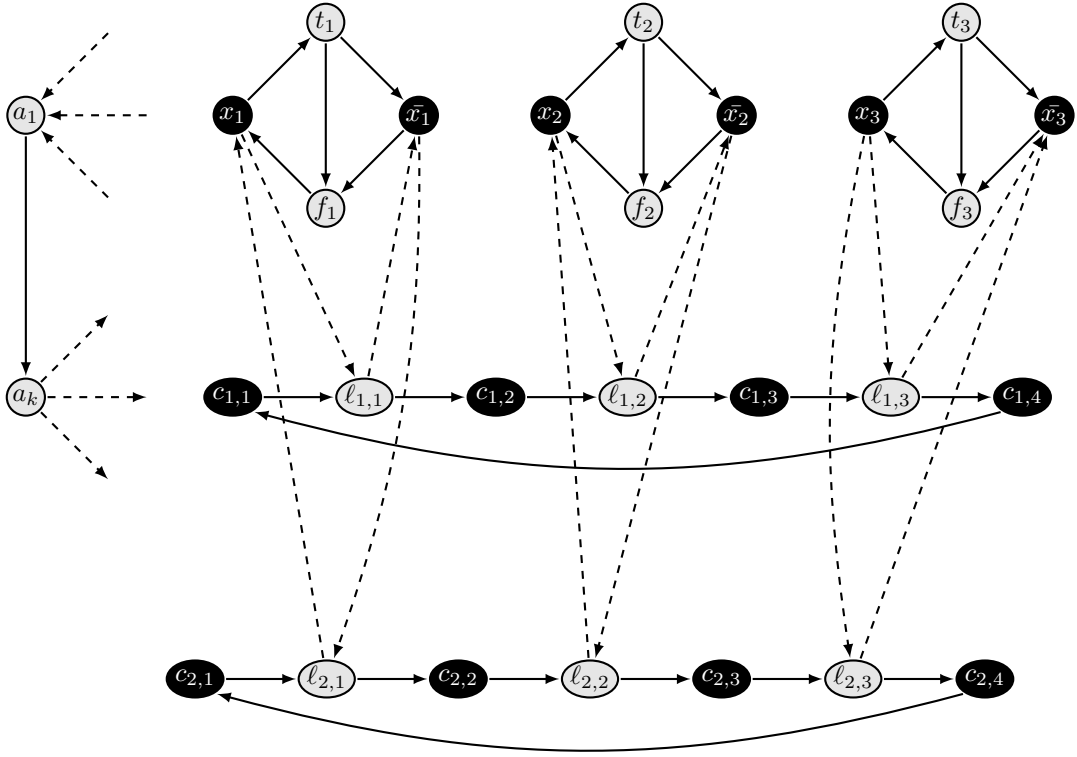
\begin{figure}[h] \label{figure:reduction_D_psi}
	\begin{center}
		\input{fig/reduction}
	\end{center}
	\caption{Representation of the digraph $D^\psi$ with $\psi$ the 3-CNF formula $\lambda_1 \vee \lambda_2 \vee \lambda_3 \wedge \neg \lambda_1 \vee \neg \lambda_2 \vee \lambda_3$.
To not put too much arcs, the out-arcs of $a_k$ and the in-arcs of $a_1$ are not represented. The vertices $v$ such that there exists an arc $(v,a_1)$ and another arc $(a_k,v)$ are fill in black.	
}
	\end{figure}
	
	Now, let us prove that if the $2$-SLCE problem has a solution for an instance, then the corresponding 3-SAT too. 
	Note that if the $2$-SLCE problem has a solution then the $2$-BLCE problem too because a sequential update schedule is a special block-sequential update schedule.

	Now, let us prove that if $\psi$ can be satisfied, there is a solution to the corresponding instance $2$-SLCE instance.
	Let $g$ be a valuation of $\lambda$ that satisfies $\psi$.
	Consider the following partition of $V$.
	In $V_1$ we put:
	\begin{itemize}
	    \item $a_1$
		\item For all variables $\lambda_i$,
		$t_i$ if $g(\lambda_i) = \top$ or $f_i$ if $g(\lambda_i) = \bot$.
		\item For all literal $\mu_{j,p}$, $\ell_{j,p}$ $g(\mu_{j,p}) = \top$ (\textit{i.e} $g(\lambda_i) = \rho$ with $\mu_{j,p} = (\lambda_i,\rho)$).
	\end{itemize}
	In $V_0$ we put the others.
	An example of partition is given in \refer{Figure}{figure:partition_D_psi}.
	

	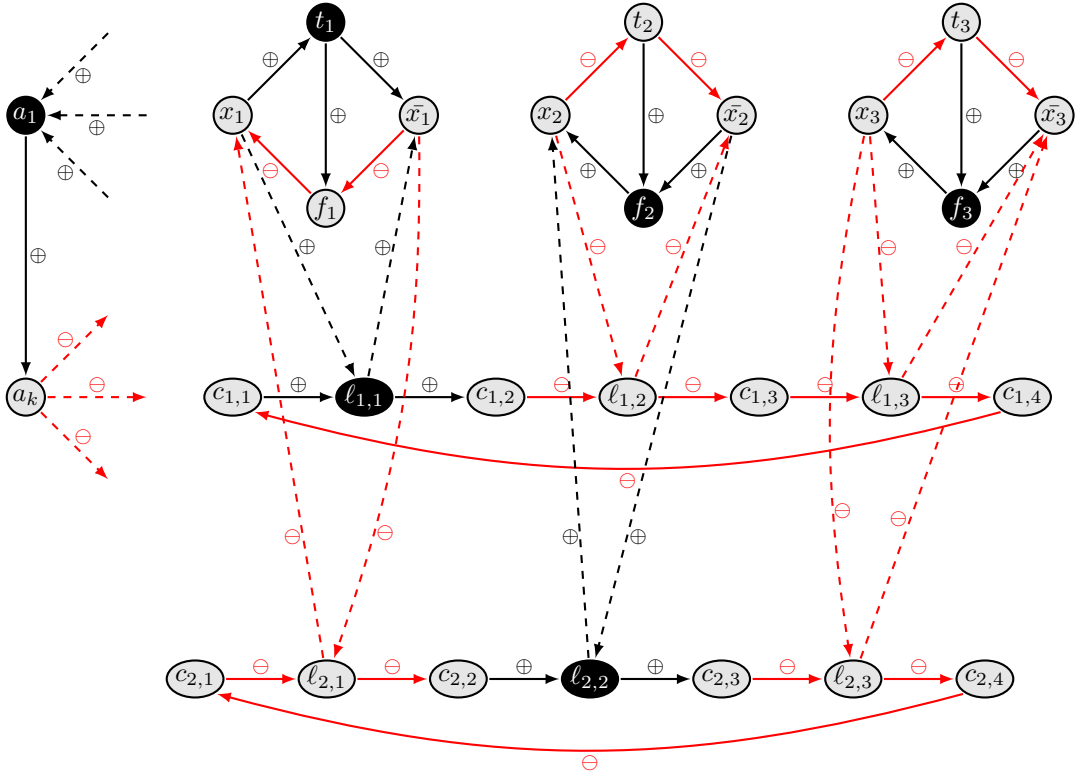
\begin{figure}[h] \label{figure:partition_D_psi}
	\begin{center}
		\input{fig/reduction_solution}
	\end{center}
	\caption{Representation of a partition of $D^\psi$ with $\psi$ the 3-CNF formula $\lambda_1 \vee \lambda_2 \vee \lambda_3 \wedge \neg \lambda_1 \vee \neg \lambda_2 \vee \lambda_3$ which make it a $2$-labeled sequential update digraph.
	In this example we take $g(\lambda_1) = \top, g(\lambda_2) = \bot$ and $g(\lambda_3) = \bot$ and we represent $V_1$ in black and $V_0$ in light gray).
	}
\end{figure}

    For all $u$, let $V(u) = i$ such that $u \in V_i$.
	Then, let $\lab$ be the labeling such that for all $(u,v) \in A$, $\lab{(u,v)} = \me$ if $V(u) = V(v)$ and $\lab{(u,v)} = \mi$ otherwise.
	To prove that this is a solution for $2$-SLCE, it is sufficient to prove that the reversed digraph $D' = D_{\lab}^{R}$, is acyclic by Lemma~\ref{lem:car}.
	
	First, one can show that there is no cycle in $D'$ passing through $a_k$ or $b$.
	Indeed, the only incoming arc of $a_k$ in $D$ is $(b,a)$ which is positive, and the outgoing arcs of $a_k$ are all negatives. 
	Therefore, $a_k$ has no outgoing arcs in $D'$ and no cycle go through $a_k$.
	Similarly, the only outgoing arcs of $a_1$ in $D'$ go to $a_k$. This signifies that no cycle can go through $b$.
	
	Second, one can show that no cycle in $D'$ passes through vertices $x_i,\bar{x_i}$ with $i \in [n]$.
	From now on, we will ignore the arcs $(a_k,x_i)$, $(\bar{x_i},a_1)$, $(a_k,\bar{x_i})$ and $(x_i,a_1)$ because we already said that no cycle go through $a_1$ or $a_k$.
	There are two cases to consider. 
	First, consider that $g(\lambda_i) = \top$.
	Then, all the out-neighbors of $x_i$ in $D$ are in $V_1$.
	Indeed, $t_i$ is in $V_1$ and the vertices $\ell_{j,p}$ such that $\mu_{j,p} = (\lambda_i,\top)$ are also in $V_1$.
	Furthermore, all in-neighbors of $x_i$  are in $V_0$.
	Indeed, $f_i$ is in $V_0$ and the vertices $\ell_{j,p}$  such that $\mu_{j,p} = (\lambda_i,\bot)$ are also in $V_0$.
	Therefore, all in-arcs of $x_i$ in $D$ are negative arcs and all out-arc are positive.
	This signifies that $x_i$ has no in-neighbors in $D'$ and therefore no cycle go through $x_i$.
	The same way, all in-arcs of $\bar{x_i}$ in $D$ are positive and all out-arc are negative arcs. 
	Therefore, $\bar{x_i}$ has no out-neighbors in $D'$ and therefore no cycle go through $\bar{x_i}$.
	The case where $g(\lambda_i) = \bot$ is totally symmetrical.
	Because no cycle go through $x_i$ and $\bar{x_i}$ in $D'$, one can see that no cycle can go through $t_i$ or $f_i$ neither.
	
	Finally, the only remaining vertices are those in $C$ and $L$.
	The graph $D_\psi$ restricted to these vertices is only composed of $m$ disjoint cycles corresponding to the $m$ clauses.
	Consider the cycle that correspond to a clause $\mu_j$.
	One can see that the arc $(c_{j,4},c_{j,1})$ is negative (because $c_{j,4}$ and $c_{j,1}$ are in $V_0$).
	Hence, for a cycle in $D'$  exists in the component corresponding to the $\mu_j$, all the arcs have to be negative and therefore, all the vertices have to be in $V_0$.
	However, since $g$ is a valid solution of $\psi$, one of the vertices $\ell_{j,1}$, $\ell_{j,2}$ or $\ell_{j,3}$ is in $V_1$.
	As a result, $D'$ has no cycles and $2$-SLCE (and therefore $2$-BLCE) has a solution.
	
	Now, let us consider a partition of $2$-BLCE on a instance $D_\psi$ and prove that there is also a solution of $2$-SLCE and that it gives a solution for the corresponding $3$-CNF.

	Conversely, suppose that there exists a block-sequential update schedule $s'$ such that $\Phi_k(f^{s'}) \neq \emptyset$.
	
	First, let us prove that there exists a block-sequential update schedule $s$ which update first a block $B_1 = \{a_k\}$ and such that $\Phi_k(f^{s}) \neq \emptyset$.
	First, by \refer{Lemma}{lem:perm}, it is possible to update the block $B_1$ which contains $a_k$ in first.
	Second, by~\cite{Math11}, if is impossible to update $a_k$ alone, then it is because there is a cycle in $D^f_s$ containing $a_k$ with only positive arcs. 
	All cycles in $D^f$ going through $a_k$ are passing through $a_1$ because $\inNeighbors_{D^f}(a_k) = \{a_1\}$.
	Furthermore, the positive cycles are necessarily going through a vertex $v \in X \cup C$ because $\outNeighbors_{D^f}(a_k) = X \cup C$.
	This means that $a_1,a_k$ and $v$ are updated in the same block $B_1$ and $s(a_k) = s(a_1) = s(v)$.
	Furthermore, because $\inNeighbors(a_1) = X \cup C$, there exists an arc $(v,a_1)$ and because $s(a_1) = s(v)$, the arc $(v,a_1)$ is positive like the arcs $(a_1,a_k)$ and $(a_k,v)$.
	As a result, the cycle $(a_1,a_k,v)$ has $3$ positive arcs, which is not a multiple of $k=2$ and $D$ is not a $k$-labeling which is a contradiction. 
	As a result, we can find a block-sequential update schedule $s=B_1, \dots, B_m$ such that $B_1 = \{a_k\}$ such that $\Phi_2(f^{s}) \neq \emptyset$ by~\cite{Math11}.
	
	Therefore, $D^s = (D,\lab)$ is a $2$-labeling and therefore there is a partition $\{V_0,V_1\}$ of $V$ such that for all $(u,v) \in A$, $\lab(u,v)$ is positive iff $V(u) \neq V(v)$.
	
	For all $u \in V$, let $V(u) = i$ such that $u \in B_i$.
	Without loss of generality, suppose $V(a_k) = 0$.
	Since, $B_1 = \{a_k\}$, $s(a_k) = 1$ then for all $v \in X \cup C$, we have $s(a_k) < s(v)$ and $(a_k,v) \in A$. Therefore, $\lab(a_k,v) = \me$ and then $V(v) = V(a_k) = 0$.
	
	Now, let us prove that for all $i \in [n]$, $V(t_i) \neq V(f_i)$.
	For the sake of contradiction, suppose that $V(t_i) = V(f_i)$.
	There are two cases. 
	\begin{itemize}
		\item If $V(t_i) = V(f_i) = 0$, then $\lab(x_i,t_i) = \lab(t_i,f_i) = \lab(f_i,x_i) = \me$ and $(x_i,f_i,t_i)$ is a cycle of $(D^f_s)^R$ with only negative arcs.
		It is a contradiction, because $D^f_s$ would not be an update digraph.
		\item If  $V(t_i) = V(f_i) = 1$ then $\lab(t_i,\overline{x_i}) = \lab(\overline{x_i},f_i) = \mi$, $\lab(t_i,f_i) = \me$ and $(\overline{x_i},f_i,t_i)$ is a cycle of $(D^f_s)^R$ with a negative $(f_i,ti)$ arc.
		It is a contradiction, because $D^f_s$ would not be an update digraph.
	\end{itemize}
	As a result, we have $V(t_i) \neq V(f_i)$ for all $i \in [n]$.
	Note that there are two cases: 
	\begin{itemize}
		\item $V(t_i) = 0$, $V(f_i) = 1$, $\lab(x_i,t_i) = \lab(t_i,\overline{x_i}) = \me$ and $\lab(\overline{x_i},f_i) = \lab(f_i,x_i) = \mi$.
		\item $V(t_i) = 1$, $V(f_i) = 0$, $\lab(x_i,t_i) = \lab(t_i,\overline{x_i}) = \mi$ and $\lab(\overline{x_i},f_i) = \lab(f_i,x_i) = \me$.
	\end{itemize}
	In these two cases, since $V(t_i) \neq V(f_i)$, $\lab(f_i,t_i) = \mi$.

	Consider the valuation $g:\lambda_i \mapsto \begin{cases} \top &\text{ if } V(t_i) = 1 \\ \bot  &\text{ if } V(t_i) = 0   \end{cases}$.
	In the following, we prove that $g$ is a valuation that satisfies $\psi$.

	Let us prove that, for all $j \in m$, $p \in [3]$, we have $g(\mu_{j,p}) = \top$ if and only if $V(\ell_{j,p}) = 1$.
	To prove that, it is sufficient to show that $V(\ell_{j,p}) = V( t_i )$ if $\mu_{j,p} = (\lambda_i, \top)$ and $V(\ell_{j,p}) = V( f_i )$ if $\mu_{j,p} = (\lambda_i, \bot)$.
	
	First, suppose that $\mu_{j,p} = (\lambda_i, \top)$. 
	Recall that wa have two arcs $(x_i,\ell_{j,p})$ and $(\ell_{j,p},\overline{x_i})$.
	Now, for the sake of contradiction, suppose that $V(\ell_{j,p}) \neq V( t_i )$.
	There are two cases.
	\begin{itemize}
		\item  If $V( t_i ) = 0$ and $V(\ell_{j,p})=1$, then  $\lab(x_i,t_i) = \lab(t_i,\overline{x_i}) = \me$ and $\lab(x_i, \ell_{j,p}) = \lab(\ell_{j,p},\overline{x_i}) = \mi$.
		As a result, $(x_i,\ell_{j,p},\overline{x_i},t_i,x_i)$ is a cycle in $(D^f_s)^R$ with two negative arcs ($(\overline{x_i},t_i)$ and $(t_i,x_i)$).
		\item  If $V( t_i ) = 1$ and $V(\ell_{j,p})=0$, then  $\lab(x_i,t_i) = \lab(t_i,\overline{x_i}) = \mi$ and $\lab(x_i, \ell_{j,p}) = \lab(\ell_{j,p},\overline{x_i}) = \me$.
		As a result, $(x_i,t_i,\overline{x_i},\ell_{j,p},x_i)$ is a cycle in $(D^f_s)^R$ with two negative arcs	($(\overline{x_i},\ell_{j,p})$ and $(t_i,x_i)$).	
	\end{itemize}
	Both cases are impossible because it would mean that $D^f_s$ is not an update digraph.
	Therefore, if $\mu_{j,p} = (\lambda_i, \top)$ then $g(\mu_{j,p}) = \top$ if and only if $V(\ell_{j,p}) = 1$.
	The case where $\mu_{j,p} = (\lambda_i, \bot)$ is totally symmetrical. 
	
	Finally, let us prove that all clauses are satisfied by $g$ (and then that the formula $\psi$ is satisfied by $g$).
	In other word, let us prove that for all $j \in [m]$ there exists $p \in [3]$ such that $g(\mu_{j,p}) = \top$.
	By contradiction, suppose that for some $j \in [m]$, for all $p \in [3]$, we have $g(\mu_{j,p}) = \bot$.
	This means that 
	\[
	V(c_{j,1}) = V(\ell_{j,1}) = V(c_{j,2}) = V(\ell_{j,2}) = V(c_{j,3}) = V(\ell_{j,3}) = V(c_{j,4}) = 0
	\] 
	and then that $(c_{j,4}, \ell_{j,3}, c_{j,3}, \ell_{j,2}, c_{j,2}, \ell_{j,1}, c_{j,1})$ is a cycle of $(D^f_s)^R$ with only negative arcs which is a contradiction.
	
	Then $g$ is a valid valuation of $\psi$.
	Furthermore, with the same reasoning as earlier, we can show that $D^f_s$ is an sequential update digraph.
	Indeed:
	\begin{itemize}
		\item No positive cycles of $D^f_s$ go through $a_k$ and then through $b$. 
		\item For any $i \in [n]$, no positive cycles of $D^f_s$ go through $x_i$ or $\overline{x_i}$ and then through $t_i$ or $f_i$.
		\item For any $j \in [m]$, no positive cycle can go through $c_{j,1}$ because $\lab(c_{j,4},c_{j,1}) = \me$  and then through $L \cup C$.
	\end{itemize} 
	
	It means that if there is a block-sequential solution, then there is a sequential one and this solution corresponds to a valuation of $\psi$.
	Hence, we finished the reduction from 3-SAT to 2-PLCE and 2-BLCE and they are NP-hard.
	
	\qed
\end{pf}


\begin{lem} \label{lem:k-BLCE_hard}
	For any $k \geq 2$, $k$-BLCE and $k$-SLCE problems are NP-hard.
\end{lem}
\begin{pf}
	
	Consider the digraph $D^{\psi,k} = (V',A')$ which is constructed from the digraph $D^\psi = (V,A)$ of \refer{Lemma}{lem:NP_hard_2}. More precisely,   the following arcs of $D^\psi$ will be replaced by isolated paths of length $k-1$ in $D^{\psi,k}$:
\begin{itemize}
	\item $(a_1,a_k)$	
\item $(t_i,\overline{x_i})$ for any $i \in [n]$
\item $(t_i,f_i)$ for any $i \in [n]$
\item $(f_i,x_i)$ for any $i \in [n]$
\item $(c_{j,p},\ell_{j,p})$ for any $j \in [m]$
\item $(c_{j,p},x_i)$ or $(c_{j,p},\overline{x_i})$ for $j \in [m]$ and $i \in [n]$ such that the arc exists.
\end{itemize}

An example of digraph $D^{\psi,k}$ for $k=3$ is given \refer{Figure}{figure:reduction_k}.

	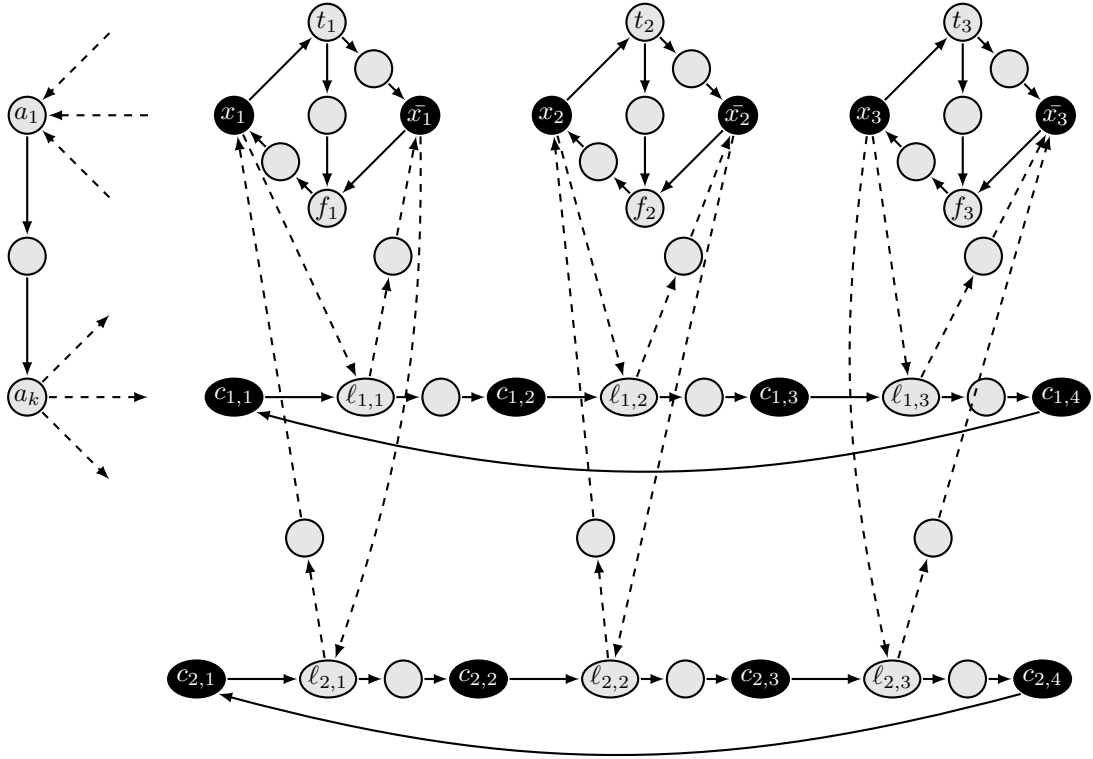
\begin{figure}[h] \label{figure:reduction_k}
	\begin{center}
		\input{fig/reduction_3}
	\end{center}
	\caption{Representation of the digraph $D^{\psi,k}$ with $\psi$ the 3-CNF formula $\lambda_1 \vee \lambda_2 \vee \lambda_3 \wedge \neg \lambda_1 \vee \neg \lambda_2 \vee \lambda_3$ and $k=3$.
	It differs from $D^\psi$ because some arcs are replace by isolated paths of length $k-1 = 2$.
	}
\end{figure}

	The proof is the same that \refer{Lemma}{lem:NP_hard_2}, but instead of the digraph $D^\psi$, we use the digraph $D' = D^{\psi,k}$.
	
	If $\psi$ has a solution then there is a labeling $\lab$ of $D^\psi$ such that $D^\psi_{\lab}$ is a $2$-labeling sequential update digraph.
	From this construction, we can construct a labeling $\lab'$ such that 
	$D'_{\lab'}$ is a $k$-labeling sequential update digraph.
	
	Consider the partition function $h:V \to [0,1]$ such that for all $\ell \in \{0,1\}$, and for all $i \in V_\ell$, we have $h(i) = \ell$.
	We define the new partition function $h': V' \to [0,k[$ as follows.
	First, for all $v \in V$, $h'(v) = h(v) \in \{0,1\}$.
	Second, for all arc $(v^1,v^k) \in A$ replaced by a path $(v^1,v^2,\dots,v^k)$ in $D'$ there are three cases:
	\begin{itemize}
		\item If $h(v^1) = h(v^k)$ (and therefore the arc $(v^1,v^k)$ is negative) then $h'(v^1) = h'(v^2) =  \dots= h'(v^k)$ (and therefore all the path is negative). 
		\item If $h(v^1) = 1$ and $h(v^k) = 0$ (and therefore, in $D_\psi$, the arc $(v^1,v^k)$ is positive), then for all $i \in [2,k]$, we have $h'(v^i) = h'(v^{i-1})+1 \bmod k$.
In other words, $h'(v^1) = 1, h'(v^2) = 2, \dots, h'(v^{k-1}) = k-1$ and $h'(v^k) = 0$ and all the path is positive in $D'$.
		\item If $h(v^1) = 0$ and $h(v^1) = 1$ (and therefore the arc $(v^1,v^k)$ is positive), then $h'(v^1) = 0$ and 
		$h'(v^2) = h'(v^3) = \dots = h'(v^k) = 1$ and the path is composed of one positive arc and $k-2$ negative arcs.
	\end{itemize}
	
	One can check that for all arc $(v,v') \in A'$ either $h'(v') = h'(v)$ (and therefore $\lab'(v,v') = \me$) or $h'(v') = (h'(v) +1) \bmod k$ (and therefore $\lab'(v,v') = \mi$).
	As a result, $D'_{\lab{}'}$ is $k$-labeled.
	Furthermore, one can see that $D'_{\lab{}'}$ is a sequential update digraph.
	Indeed, as seen in \refer{Lemma}{lem:NP_hard_2}, $D_{\lab{}}$ is a sequential update digraph and therefore $D_{\lab{}}^R$ is acyclic.
	Moreover, there are few differences between ${D}_{\lab{}}^R$ and ${D'}_{\lab{}'}^R$:
	some arcs $(v^1,v^k)$ of $D_{\lab{}}$ are replaced by a path $(v^1,v^2,\dots,v^k)$ in $D'_{\lab{}'}$ .
	However, as we have seen, if $\lab(v^1,v^k) = \mi$ then the entire path  $(v^1,v^2,\dots,v^k)$ is positive in ${D'}_{\lab{}'}$ and will therefore not add any cycle in ${D'}_{\lab{}'}^R$.
	The same way, if $\lab(v^1,v^k) = \mi$ then either the entire path  $(v^1,v^2,\dots,v^k)$ is negative (and then the negative arc $(v^k,v^1)$ of $D_{\lab{}}^R$ is replaced by a negative path in ${D'}_{\lab{}'}^R$), or $(v^1,v^2)$ is positive and all the path $(v^2,\dots,v^k)$ is negative (and then the negative arc $(v^k,v^1)$ of $D_{\lab{}}^R$ is replace by the two paths $(v^1,v^2)$ and $(v^k,\dots,v^2)$ in ${D'}_{\lab{}'}^R$). 
	In both cases, there are no possible cycles added. 
	Therefore, ${D'}_{\lab{}'}^R$ is acyclic and ${D'}_{\lab{}'}$ is a sequential update digraph.
	
	Conversely, consider a block-sequential update schedule $s'$ such that of $D'_{s'}$ is an update digraph. 
	One can prove, similarly that in \refer{Lemma}{lem:NP_hard_2} that there exists another block-sequential $s$ which updates $\{a_k\}$ first.
	Indeed, from \refer{Lemma}{lem:perm}, we know that we can update the block containing $a_k$ first.
	Furthermore, we can prove that we can update $a_k$ only in the first block because if $a_k$ is in a positive cycle then it is in a positive cycle of length $k+1$ which would contradict the fact that $D'_{s}$ is a $k$ labeling.
	
	As a result, for all $v \in X \cup C$, we have $h'(v) = h'(a_k) = 0$.
	On the other hands, for all $v' \in V$ there exists $v \in X \cup C$ such that $(v,v') \in A'$.
	Therefore, for all $v \in V$, we have $h'(v) \in \{0,1\}$.
	
	Now, we prove that for all $i \in [n]$, $h'(t_i) \neq h'(f_i)$.
	For the sake of contradiction, let us suppose that $h'(t_i) = h'(f_i)$. 
	Note that the path between $t_i$ and $f_i$ is only composed of negative arcs because there are only $k-1$ arcs between $t_i$ and $f_i$ and $h'(t_i) = h'(f_i)$.
	There are two cases.
	First, if $h'(t_i) = h'(f_i) = 0$ then the path $(f_i,\dots, x_i)$ and the arc $(x_i,t_i)$ are full negative because $h'(x_i) = h'(t_i) = h'(f_i) = 0$.
	This means that there is a cycle $(t_i, x_i, \dots, f_i, \dots, t_i)$ in ${D'}^R_{\lab'}$ with only negative arcs.
	Second, if $h'(t_i) = h'(f_i) = 1$ then the path $(t_i,\dots, \bar{x_i})$ and the path $(\bar{x_i},f_i)$ are positive $h'(\bar{x_i})$ and $h'(t_i) = h'(f_i) = 1$.
	This means that there is a cycle $(t_i, \dots, \bar{x_i}, f_i, \dots, t_i)$ in ${D'}^R_{\lab'}$ with negative arcs in the path $(f_i,\dots,t_i)$.
	As a result, we have $h'(t_i) \neq h'(f_i)$.
	
	Consider the valuation $g:\lambda_i \mapsto \begin{cases} \top &\text{ if } h'(t_i) = 1 \\ \bot  &\text{ if } h'(t_i) = 0   \end{cases}$.
	In the following, we prove that $g$ is a valid valuation of $\psi$.

	Let us prove that, for all $j \in m$, $p \in [3]$, we have $g(\mu_{j,p}) = \top$ if and only if $h(\ell_{j,p}) = 1$.
	
	To prove that, it is sufficient to show that $h'(\ell_{j,p}) = h'( t_i )$ if $\mu_{j,p} = (\lambda_i, \top)$ and $h'(\ell_{j,p}) = h'( f_i )$ if $\mu_{j,p} = (\lambda_i, \bot)$.
	
	First, suppose that $\mu_{j,p} = (\lambda_i, \top)$. 
	Recall that wa have an arc $(x_i,\ell_{j,p})$ and a path $(\ell_{j,p},\dots,\overline{x_i})$.
	Now, for the sake of contradiction, suppose that $h(\ell_{j,p}) \neq h( t_i )$.
	There are two cases.
	\begin{itemize}
		\item  If $h'( t_i ) = 0$ and $h'(\ell_{j,p})=1$, then
		the path $(x_i,t_i, \dots, \overline{x_i})$ is fully negative when the path $(x_i,\ell_{j,p},\dots,\overline{x_i})$ is fully positive.
		As a result, $(x_i,\ell_{j,p},\dots,\overline{x_i},\dots,t_i,x_i)$ is a cycle in $(D')^R_{\lab'}$ with negative arcs.
		\item If $h'( t_i ) = 1$ and $h'(\ell_{j,p})=0$, then
		the path $(x_i,t_i, \dots, \overline{x_i})$ is fully positive when the path $(x_i,\ell_{j,p}, \dots, \overline{x_i})$ is fully negative.
		As a result, $(x_i,t_i,\dots,\overline{x_i},\dots,\ell_{j,p},x_i)$ is a cycle in $(D')^R_{\lab'}$ with negative arcs.
	\end{itemize}	Both cases are impossible because it would mean that $D'_{\lab'}$ is not an update digraph and then if $\mu_{j,p} = (\lambda_i, \top)$ then $g(\mu_{j,p}) = \top$ if and only if $h'(\ell_{j,p}) = 1$.
	The $\mu_{j,p} = (\lambda_i, \bot)$ case is totally symmetrical. 
	
	Finally, let us prove that all clauses are satisfied by $g$ (and then that the formula $\psi$ is satisfied by $g$).
	In other word, let us prove that for all $j \in [m]$ there exists $p \in [3]$ such that $g(\mu_{j,p}) = \top$.
	By contradiction, suppose that for some $j \in [m]$, for all $p \in [3]$, we have $g(\mu_{j,p}) = \bot$.
	This signifies that 
	\[
	h(c_{j,1}) = h(\ell_{j,1}) = h(c_{j,2}) = h(\ell_{j,2}) = h(c_{j,3}) = h(\ell_{j,3}) = h(c_{j,4}) = 0
	\] and therefore that all paths $(\ell_{j,p},\dots,c_{j,p+1})$ are fully negative
	and then that $(c_{j,4}, \dots, \ell_{j,3}, c_{j,3}, \dots, \ell_{j,2}, c_{j,2}, \dots, \ell_{j,1}, c_{j,1})$ is a cycle of $(D')^R$ with only negative arcs which is a contradiction.
	
	Then $g$ is a valid valuation of $\psi$.
	Furthermore, with the same reasoning than earlier, we can show that $D'$ is an sequential update digraph.
	Indeed:
	\begin{itemize}
		\item No positive cycles of $D'$ go through $a_k$ and then through $b$. 
		\item For any $i \in [n]$, no positive cycles of $D'$ go through $x_i$ or $\overline{x_i}$ and then through $t_i$ or $f_i$.
		\item For any $j \in [m]$, no positive cycle can go through $c_{j,1}$ because $\lab(c_{j,4},c_{j,1}) = \me$  and then through $L \cup C$.
	\end{itemize} 
	
	It means that if there is a block-sequential solution, then there is a sequential one and this solution corresponds to a valuation of $\psi$.
	Hence, we finished the reduction from 3-SAT to k-PLCE and k-BLCE and they are NP-hard.
	\qed
\end{pf}

\section{Acknowledgments}
Julio Aracena was supported by ANID–Chile through Centro de Modelamiento Matem\'atico (CMM), ACE210010 and FB210005, BASAL funds for center of excellence from ANID-Chile. Julio Aracena, Luis G\'omez and Lilian Salinas were supported by Conicyt–Chile through Fondecyt 1131013. Julio Aracena, Lilian Salinas and Florian Bridoux were supported by ANID–Chile through ECOS C19E02. Florian Bridoux was supported by the Young Researcher project ANR-18-CE40-0002-01 “FANs”.
\section{Conclusion}

In this paper, we study the complexity of the problem of determining if a conjunctive network has a limit cycle of a given length $k$.

In a first part, we study this problem with a parallel update schedule. We show that the problem is in P when the interaction digraph of the conjunctive network is strongly connected.
Furthermore, we also prove that the problem is NP-complete when $k$ is a parameter of the problem and the interaction digraph is not strongly connected.
However, the case where the interaction digraph is not strongly connected, but $k$ is fixed remains open. 
As a side result, we proved that the lengths of the limit cycle of a conjunctive network of length $n$ cannot divide a prime greater than $n$ and that the maximum length of a limit cycle of a conjunctive network of length $n$ corresponds to the Landau's function $g(n)$.

In a second part, we study this problem with block-sequential and sequential update schedules.
The problem is then: is there a block-sequential (\textit{resp.} a sequential) update schedule $s$ such that $f^s$ has a limit cycle of length $k$.
We prove that this problem, even with $k$ fixed and with sequential or block-sequential update schedule is NP-complete for all $k \geq 2$. 

\newpage

\bibliographystyle{elsarticle-harv}
\bibliography{main}
\end{document}

%% file: fig/FigEjID.tex
\begin{tabular}{cc@{\quad\quad}c}
\begin{tikzpicture}
 \foreach  \x in {1,...,4}{
	\coordinate (c\x) at (-45+90*\x:\radio);
	\node[main node](\x) at (c\x) {$\x$};
 }
 \Loop{1}{45}{};
 \path[arcos]
 (1) edge[bend right=10] (2)
 (2) edge[bend right=10] (3)
 (3) edge[bend right=10] (4)
 (4) edge (2)
 (4) edge[bend right=10] (1);
\end{tikzpicture}
&
$\begin{array}[b]{l}
 f_1(x)= x_1\land x_4\\
 f_2(x)= x_1\lor x_4\\
 f_3(x)= x_2\\
 f_4(x)= x_3\\
 \end{array}
$
&
\begin{tikzpicture}
 \foreach  \x in {1,...,4}{
	\node[main node](\x) at (c\x) {$\x$};
 }
 \Loop{1}{45}{$\oplus$};
 \path[arcos]
 (1) edge[bend right=10] node[labels] {$\ominus$} (2)
 (2) edge[bend right=10] node[labels] {$\ominus$} (3)
 (3) edge[bend right=10] node[labels] {$\ominus$} (4)
 (4) edge node[labels] {$\oplus$} (2)
 (4) edge[bend right=10] node[labels] {$\oplus$} (1);
\end{tikzpicture}
\\
&&$s=(1,2,3,4)$\\
\multicolumn{2}{c}{a)}&b)\\
\end{tabular}

%% file: fig/PGFs.tex
\arraycolsep=1pt
\begin{tabular}{ccc}
$D^{f}_{s}$&$\mathcal{P}\paren{D^{f}_s}$&$D^{f^{s}}$\\
\begin{tikzpicture}[baseline=(3)]
 \foreach  \x in {1,...,4}{
	\coordinate (c\x) at (-45+90*\x:\radio);
	\node[main node](\x) at (c\x) {$\x$};
 }
 \Loop{1}{45}{$\oplus$};
 \path[arcos]
 (1) edge node[labels] {$\ominus$}  (2)
 (2) edge node[labels] {$\ominus$} (3)
 (3) edge node[labels] {$\ominus$} (4)
 (4) edge node[labels] {$\oplus$} (2)
 (4) edge node[labels] {$\oplus$} (1)
 ;
\end{tikzpicture}&
\begin{tikzpicture}[baseline=(3)]
 \foreach  \x in {1,...,4}{
	\node[main node](\x) at (c\x) {$\x$};
 }
 \Loop{1}{45}{};
 \Loop{4}{-45}{};
 \path[arcos]
 (1) edge (2)
 (1) edge (3)
 (1) edge[bend right=15] (4)
 (4) edge (2)
 (4) edge (3)
 (4) edge[bend right=15] (1)
 ;
\end{tikzpicture}&
\begin{tikzpicture}[baseline=(3)]
 \foreach  \x in {1,...,4}{
	\node[main node](\x) at (c\x) {$\x$};
 }
 \Loop{1}{45}{};
 \Loop{4}{-45}{};
 \path[arcos]
 (4) edge (2)
 (4) edge (3)
 (4) edge (1)
 ;
\end{tikzpicture}\\
$\begin{array}{rl}
 f_1(x)&= x_1\land x_4\\
 f_2(x)&= x_1\lor x_4\\
 f_3(x)&= x_2\\
 f_4(x)&= x_3\\ 
s&=(1,2,3,4)
 \end{array}$&
&
$\begin{array}{rl}
 f_1^{s}(x)&= x_1\land x_4\\
 f_2^{s}(x)&= x_4\\
 f_3^{s}(x)&= x_4\\
 f_4^{s}(x)&= x_4
 \end{array}$
\end{tabular}

%% file: fig/FigPrimitive.tex
\begin{tabular}{ccl}
$G_{\lab}$& $\mathcal{P}(G_{\lab})$& \multicolumn{1}{c}{Dynamics}\\
\begin{tikzpicture}
 \foreach \x in {1,...,4}{
	\coordinate (c\x) at (-45+\x*90:0.8\radio);
	\node[main node] (\x) at (c\x) {$\x$};
 }
 \path[arcos]
 (1) edge node[labels] {$\mi$} (2)
 (2) edge node[labels] {$\mi$} (3)
 (3) edge node[labels] {$\mi$} (1)
 (1) edge[bend left =10] node[labels] {$\mi$} (4)
 (4) edge[bend left =10] node[labels] {$\mi$} (1)
 ;
\end{tikzpicture}&
\begin{tikzpicture}
 \foreach \x in {1,...,4}{
	\coordinate (c\x) at (-45+\x*90:0.8\radio);
	\node[main node] (\x) at (c\x) {$\x$};
 }
 \path[arcos]
 (1) edge  (2)
 (2) edge  (3)
 (3) edge  (1)
 (1) edge[bend left =10]  (4)
 (4) edge[bend left =10]  (1)
 ;
\end{tikzpicture}&
\begin{tikzpicture}
\node[tipo] (0) at (180:1) {};
\node[tipo] (15) at (0.00:0) {};
\node[tipo] (14) at (-20.00:1) {};
\node[tipo] (13) at (0:1) {};
\node[tipo, shift={(-20:1)}] (11) at (13) {};
\node[tipo, shift={(20:1)}] (9) at (13) {};
\node[tipo] (10) at (0:2) {};
\node[tipo] (7) at (-6.67:3) {};
\node[tipo, shift={(0:1)}] (12) at (7) {};
\node[tipo] (6) at (6.67:3) {};
\node[tipo] (5) at (0:3) {};
\node[tipo] (8) at (0:4) {};
\node[tipo, shift={(-20:1)}] (3) at (8) {};
\node[tipo, shift={(0:1)}] (2) at (8) {};
\node[tipo, shift={(20:1)}] (1) at (8) {};
\node[tipo, shift={(0:1)}] (4) at (2) {};
\path[-to]
(0) edge[loop left] (0)
(15) edge[loop left] (15)
(14) edge (15)
(13) edge (15)
(11) edge (13)
(10) edge (13)
(7) edge (10)
(12) edge (7)
(6) edge (10)
(5) edge (10)
(8) edge (5)
(3) edge (8)
(2) edge (8)
(4) edge (2)
(1) edge (8)
(9) edge (13);
\end{tikzpicture}
\\
\begin{tikzpicture}
 \foreach \x in {1,...,4}{
	\node[main node] (\x) at (c\x) {$\x$};
 }
 \path[arcos]
 (1) edge node[labels] {$\me$} (2)
 (2) edge node[labels] {$\mi$} (3)
 (3) edge node[labels] {$\mi$} (1)
 (1) edge[bend left =10] node[labels] {$\mi$} (4)
 (4) edge[bend left =10] node[labels] {$\mi$} (1)
 ;
\end{tikzpicture}&
\begin{tikzpicture}
 \foreach \x in {1,...,4}{
	\node[main node] (\x) at (c\x) {$\x$};
 }
 \path[arcos]
 (2) edge[bend left =10]  (3)
 (3) edge (1)
 (1) edge[bend left =10] (4)
 (4) edge[bend left =10] (1)
 (3) edge[bend left =10]  (2)
 (4) edge (2)
 ;
\end{tikzpicture}&
\begin{tikzpicture}
\node[tipo] (0) at (180:1) {};
\node[tipo] (15) at (0.00:0) {};
\node[tipo] (14) at (-15.00:1) {};
\node[tipo] (13) at (15.00:1) {};
\node[tipo] (7) at (-24.00:2) {};
\node[tipo] (6) at (-15.00:2) {};
\node[tipo] (5) at (-6.00:2) {};
\node[tipo] (11) at (6.00:2) {};
\node[tipo] (10) at (15.00:2) {};
\node[tipo] (9) at (24.00:2) {};
\path[-to]
(0) edge[loop left] (0)
(15) edge[loop left] (15)
(14) edge (15)
(7) edge (14)
(6) edge (14)
(5) edge (14)
(13) edge (15)
(11) edge (13)
(10) edge (13)
(9) edge (13);
\pgftransformxshift{4cm}
\node[tipo] (12) at (0,0) {};
\node[tipo] (3) at (180.00:1) {};
\node[tipo] (2) at (-15:1) {};
\node[tipo] (4) at (-15:2) {};
\node[tipo] (1) at (15:1) {};
\node[tipo] (8) at (15:2) {};
\path[-to]
(12) edge[bend right=10] (3)
(2) edge (12)
(4) edge (2)
(1) edge (12)
(8) edge (1)
(3) edge[bend right=10] (12);
\end{tikzpicture}
\end{tabular}

%% file: FigEjRevD.tex
\begin{tabular}{cc}
\begin{tikzpicture}
 \foreach  \x in {1,...,4}{
	\coordinate (c\x) at (-45+90*\x:\radio);
	\node[main node](\x) at (c\x) {$\x$};
 }
 \Loop{1}{45}{$\oplus$};
 \path[arcos]
 (2) edge[bend right=10] node[labels] {$\ominus$} (3)
 (3) edge[bend right=10] node[labels] {$\ominus$} (4)
 (1) edge[bend left=10] node[labels] {$\ominus$} (4)
 (4) edge node[labels] {$\oplus$} (2)
 (4) edge[bend left=10] node[labels] {$\oplus$} (1);
\end{tikzpicture}&
\begin{tikzpicture}
 \foreach  \x in {1,...,4}{
	\node[main node](\x) at (c\x) {$\x$};
 }
 \Loop{1}{45}{$\oplus$};
 \path[arcos]
 (3) edge[bend left=10] node[labels] {$\ominus$} (2)
 (4) edge[bend left=10] node[labels] {$\ominus$} (3)
 (4) edge[bend right=10] node[labels] {$\ominus$} (1)
 (4) edge node[labels] {$\oplus$} (2);
\end{tikzpicture}\\
a)&b)
\end{tabular}

%% file: PGlabyGlabOrdered.tex
\begin{tabular}{cc}
\begin{tikzpicture}[scale=0.9]
 \node[main node] (1) at (0,0){1};
 \node[main node] (2) at (2.5,1){2};
 \node[main node] (7) at (2.5,-1){7};
 \node[main node] (3) at (5,3){3};
 \node[main node] (5) at (4.5,-1){5};
 \node[main node] (6) at (5.5,1){6};
 \node[main node] (8) at (5,-3){8};
 \node[main node] (4) at (7.5,0){4};
 \path[arcos]
 (1) edge node[labels] {$\oplus$} (2)
 (1) edge node[labels] {$\oplus$} (7)
 (2) edge node[labels] {$\oplus$} (3)
 (2) edge node[labels,pos=0.6] {$\oplus$} (5)
 (3) edge node[labels] {$\ominus$} (5)
 (3) edge node[labels] {$\ominus$} (6)
 (5) edge node[labels,pos=0.7] {$\oplus$} (4)
 (5) edge node[labels] {$\ominus$} (8)
 (6) edge node[labels,pos=0.7] {$\ominus$} (8)
 (6) edge node[labels] {$\ominus$} (5)
 (7) edge node[labels,pos=0.6] {$\oplus$} (3)
 (8) edge node[labels,pos=0.6] {$\oplus$} (4)
 ;
 \draw[arcos](4) to[out=90,in=0] (3.75,3.75) node[labels,above]{$\oplus$} to[out=180,in=90] (1);
 \path[draw, rounded corners, red, very thick]
 (-0.5,-1) rectangle +(1,2)
 (7)++(-0.5,-1) rectangle +(1,4)
 (8)++(-1,-1) rectangle +(2,8)
 (4)++(-0.5,-1) rectangle +(1,2)
 ;
 \node[shift={(0,-1.3)}] (l1) at (1){\color{red}$V_1$};
 \node[shift={(0,-1.3)}] (l1) at (7){\color{red}$V_2$};
 \node[shift={(0,-1.3)}] (l1) at (8){\color{red}$V_3$};
 \node[shift={(0,-1.3)}] (l1) at (4){\color{red}$V_4$};
\end{tikzpicture}
&
\begin{tikzpicture}[scale=0.9]
 \node[main node] (1) at (0,0){1};
 \node[main node] (2) at (2.5,1){2};
 \node[main node] (7) at (2.5,-1){7};
 \node[main node] (3) at (5,3){3};
 \node[main node] (5) at (4.5,-1){5};
 \node[main node] (6) at (5.5,1){6};
 \node[main node] (8) at (5,-3){8};
 \node[main node] (4) at (7.5,0){4};
 \path[arcos]
 (1) edge node[labels] {$\oplus$} (2)
 (1) edge node[labels] {$\oplus$} (7)
 (2) edge node[labels] {$\oplus$} (3)
 (2) edge node[labels,pos=0.6] {$\oplus$} (5)
 (3) edge node[labels] {$\ominus$} (5)
 (3) edge node[labels] {$\ominus$} (6)
 (5) edge node[labels,pos=0.7] {$\ominus$} (4)
 (5) edge node[labels] {$\ominus$} (8)
 (6) edge node[labels,pos=0.7] {$\ominus$} (8)
 (6) edge node[labels] {$\ominus$} (5)
 (7) edge node[labels,pos=0.6] {$\oplus$} (3)
 (8) edge node[labels,pos=0.6] {$\ominus$} (4)
 ;
 \draw[arcos](4) to[out=90,in=0] (3.75,3.75) node[labels,above]{$\oplus$} to[out=180,in=90] (1);
 \path[draw, rounded corners, red, very thick]
 (-0.5,-1) rectangle +(1,2)
 (7)++(-0.5,-1) rectangle +(1,4)
 (8)++(-1,-1) rectangle +(4,8)
 ;
 \node[shift={(0,-1.3)}] (l1) at (1){\color{red}$V_1$};
 \node[shift={(0,-1.3)}] (l1) at (7){\color{red}$V_2$};
 \node[shift={(1,-1.3)}] (l1) at (8){\color{red}$V_3$};
\end{tikzpicture}
\end{tabular}

%% file: fig/BLCEnoSLCE.tex
\begin{tikzpicture}[node distance=2cm]
\setlength{\radio}{5cm}
\node[main node] (x1) {$1$};
\node[main node, right of=x1] (x2) {$2$};
\node[main node, below of=x2] (x3) {$3$};
\node[main node, below of=x1] (x4) {$4$};
\node[main node, below right of=x1,xshift=-0.4cm,yshift=0.4cm] (x5) {$5$};

\path[arcos]
(x1) edge (x2)
(x2) edge (x3)
(x3) edge (x4)
(x4) edge (x1)
(x1) edge (x5)
(x5) edge (x2)
(x3) edge (x5)
(x5) edge (x4)
;
\end{tikzpicture}

%% file: fig/variable.tex
\begin{tikzpicture}[node distance=2cm]
\setlength{\radio}{5cm}
\node[main node] (xi) {$x_i$};
\node[main node,below right of=xi] (fi) {$f_i$};
\node[main node,above right of=xi] (ti) {$t_i$};
\node[main node,below right of=ti] (bi) {$\bar{x_i}$};
\path[arcos]
(bi) edge (fi)
(fi) edge (xi)
(xi) edge (ti)
(ti) edge (bi)
(ti) edge (fi)
;
\end{tikzpicture}

%% file: fig/clause.tex
\begin{tikzpicture}[node distance=2cm]
\setlength{\radio}{5cm}

\node[main node] (1) {$c_{j,1}$};
\node[main node,right of=1] (2) {$\ell_{j,1}$};
\node[main node,right of=2] (3) {$c_{j,2}$};
\node[main node,right of=3] (4) {$\ell_{j,2}$};
\node[main node,right of=4] (5) {$c_{j,3}$};
\node[main node,right of=5] (6) {$\ell_{j,3}$};
\node[main node,right of=6] (7) {$c_{j,4}$};

\node[above left of=2] (inv21) {};
\node[above right of=2] (inv22) {};

\node[above left of=4] (inv41) {};
\node[above right of=4] (inv42) {};

\node[above left of=6] (inv61) {};
\node[above right of=6] (inv62) {};

\path[arcos]
(1) edge (2)
(2) edge (3)
(3) edge (4)
(4) edge (5)
(5) edge (6)
(6) edge (7)
(7.south west) edge[bend left=15] (1.south east)
(inv21) edge[dashed] (2)
(2) edge[dashed] (inv22)
(inv41) edge[dashed] (4)
(4) edge[dashed] (inv42)
(inv61) edge[dashed] (6)
(6) edge[dashed] (inv62)
;
\end{tikzpicture}

%% file: fig/reduction.tex
\begin{tikzpicture}[node distance=1.75cm]
\setlength{\radio}{5cm}

\node[second node] (xi1) {$x_1$};
\node[main node,below right of=xi1] (fi1) {$f_1$};
\node[main node,above right of=xi1] (ti1) {$t_1$};
\node[second node,below right of=ti1] (bi1) {$\bar{x_1}$};
\path[arcos]
(bi1) edge (fi1)
(fi1) edge (xi1)
(xi1) edge (ti1)
(ti1) edge (bi1)
(ti1) edge (fi1)
;

\node[second node,right of=bi1,] (xi2) {$x_2$};
\node[main node,below right of=xi2] (fi2) {$f_2$};
\node[main node,above right of=xi2] (ti2) {$t_2$};
\node[second node,below right of=ti2] (bi2) {$\bar{x_2}$};
\path[arcos]
(bi2) edge (fi2)
(fi2) edge (xi2)
(xi2) edge (ti2)
(ti2) edge (bi2)
(ti2) edge (fi2)
;

\node[second node,right of=bi2] (xi3) {$x_3$};
\node[main node,below right of=xi3] (fi3) {$f_3$};
\node[main node,above right of=xi3] (ti3) {$t_3$};
\node[second node,below right of=ti3] (bi3) {$\bar{x_3}$};
\path[arcos]
(bi3) edge (fi3)
(fi3) edge (xi3)
(xi3) edge (ti3)
(ti3) edge (bi3)
(ti3) edge (fi3)
;

\node[second node,below of=xi1,yshift=-2cm] (c11) {$c_{1,1}$};
\node[main node,right of=c11] (l11) {$\ell_{1,1}$};
\node[second node,right of=l11] (c12) {$c_{1,2}$};
\node[main node,right of=c12] (l12) {$\ell_{1,2}$};
\node[second node,right of=l12] (c13) {$c_{1,3}$};
\node[main node,right of=c13] (l13) {$\ell_{1,3}$};
\node[second node,right of=l13] (c14) {$c_{1,4}$};

\path[arcos]
(c11) edge (l11)
(l11) edge (c12)
(c12) edge (l12)
(l12) edge (c13)
(c13) edge (l13)
(l13) edge (c14)
(c14.south west) edge[bend left=15] (c11.south east)

(xi1) edge[dashed] (l11)
(l11) edge[dashed] (bi1)
(xi2) edge[dashed] (l12)
(l12) edge[dashed] (bi2)
(xi3) edge[dashed] (l13)
(l13) edge[dashed] (bi3)
;

\node[second node,below of=c11,xshift=-0.5cm,yshift=-2cm] (c21) {$c_{2,1}$};
\node[main node,right of=c21] (l21) {$\ell_{2,1}$};
\node[second node,right of=l21] (c22) {$c_{2,2}$};
\node[main node,right of=c22] (l22) {$\ell_{2,2}$};
\node[second node,right of=l22] (c23) {$c_{2,3}$};
\node[main node,right of=c23] (l23) {$\ell_{2,3}$};
\node[second node,right of=l23] (c24) {$c_{2,4}$};

\path[arcos]
(c21) edge (l21)
(l21) edge (c22)
(c22) edge (l22)
(l22) edge (c23)
(c23) edge (l23)
(l23) edge (c24)
(c24.south west) edge[bend left=15] (c21.south east)

(bi1) edge[dashed,bend left=10] (l21)
(l21) edge[dashed] (xi1)
(bi2) edge[dashed] (l22)
(l22) edge[dashed] (xi2)
(xi3) edge[dashed,bend right=10] (l23)
(l23) edge[dashed] (bi3)
;

\node[main node,left of=xi1,xshift=-1cm] (b) {$a_1$};
\node[main node,below of=b,yshift=-2cm] (a) {$a_k$};
\node[above right of=b] (invb1) {};
\node[right of=b] (invb2) {};
\node[below right of=b] (invb3) {};

\node[above right of=a] (inva1) {};
\node[right of=a] (inva2) {};
\node[below right of=a] (inva3) {};

\path[arcos]
(b) edge (a)
(invb1) edge[dashed] (b)
(invb2) edge[dashed] (b)
(invb3) edge[dashed] (b)
(a) edge[dashed] (inva1)
(a) edge[dashed] (inva2)
(a) edge[dashed] (inva3)
;

\end{tikzpicture}

%% file: fig/reduction_solution.tex
\begin{tikzpicture}[node distance=1.75cm]
\setlength{\radio}{5cm}

\node[main node] (xi1) {$x_1$};
\node[main node,below right of=xi1] (fi1) {$f_1$};
\node[second node,above right of=xi1] (ti1) {$t_1$};
\node[main node,below right of=ti1] (bi1) {$\bar{x_1}$};
\path[arcos]
(bi1) edge[red] node[labels] {$\ominus$} (fi1)
(fi1) edge[red] node[labels] {$\ominus$} (xi1)
(xi1) edge node[labels] {$\oplus$} (ti1)
(ti1) edge node[labels] {$\oplus$} (bi1)
(ti1) edge node[labels] {$\oplus$} (fi1)
;

\node[main node,right of=bi1] (xi2) {$x_2$};
\node[second node,below right of=xi2] (fi2) {$f_2$};
\node[main node,above right of=xi2] (ti2) {$t_2$};
\node[main node,below right of=ti2] (bi2) {$\bar{x_2}$};
\path[arcos]
(bi2) edge node[labels] {$\oplus$} (fi2)
(fi2) edge node[labels] {$\oplus$} (xi2)
(xi2) edge[red] node[labels] {$\ominus$} (ti2)
(ti2) edge[red] node[labels] {$\ominus$} (bi2)
(ti2) edge node[labels] {$\oplus$} (fi2)
;

\node[main node,right of=bi2] (xi3) {$x_3$};
\node[second node,below right of=xi3] (fi3) {$f_3$};
\node[main node,above right of=xi3] (ti3) {$t_3$};
\node[main node,below right of=ti3] (bi3) {$\bar{x_3}$};
\path[arcos]
(bi3) edge node[labels] {$\oplus$} (fi3)
(fi3) edge node[labels] {$\oplus$} (xi3)
(xi3) edge[red] node[labels] {$\ominus$} (ti3)
(ti3) edge[red] node[labels] {$\ominus$} (bi3)
(ti3) edge node[labels] {$\oplus$} (fi3)
;

\node[main node,below of=xi1,yshift=-2cm] (c11) {$c_{1,1}$};
\node[second node,right of=c11] (l11) {$\ell_{1,1}$};
\node[main node,right of=l11] (c12) {$c_{1,2}$};
\node[main node,right of=c12] (l12) {$\ell_{1,2}$};
\node[main node,right of=l12] (c13) {$c_{1,3}$};
\node[main node,right of=c13] (l13) {$\ell_{1,3}$};
\node[main node,right of=l13] (c14) {$c_{1,4}$};

\path[arcos]
(c11) edge node[labels] {$\oplus$} (l11)
(l11) edge node[labels] {$\oplus$} (c12)
(c12) edge[red] node[labels] {$\ominus$} (l12)
(l12) edge[red] node[labels] {$\ominus$} (c13)
(c13) edge[red] node[labels] {$\ominus$} (l13)
(l13) edge[red] node[labels] {$\ominus$} (c14)
(c14.south west) edge[red,bend left=15] node[labels] {$\ominus$} (c11.south east)

(xi1) edge[dashed] node[labels] {$\oplus$} (l11)
(l11) edge[dashed] node[labels] {$\oplus$} (bi1)
(xi2) edge[dashed,red] node[labels] {$\ominus$} (l12)
(l12) edge[dashed,red] node[labels] {$\ominus$} (bi2)
(xi3) edge[dashed,red] node[labels] {$\ominus$} (l13)
(l13) edge[dashed,red] node[labels] {$\ominus$} (bi3)
;

\node[main node,below of=c11,xshift=-0.5cm,yshift=-2cm] (c21) {$c_{2,1}$};
\node[main node,right of=c21] (l21) {$\ell_{2,1}$};
\node[main node,right of=l21] (c22) {$c_{2,2}$};
\node[second node,right of=c22] (l22) {$\ell_{2,2}$};
\node[main node,right of=l22] (c23) {$c_{2,3}$};
\node[main node,right of=c23] (l23) {$\ell_{2,3}$};
\node[main node,right of=l23] (c24) {$c_{2,4}$};

\path[arcos]
(c21) edge[red] node[labels] {$\ominus$} (l21)
(l21) edge[red] node[labels] {$\ominus$}  (c22)
(c22) edge node[labels] {$\oplus$} (l22)
(l22) edge node[labels] {$\oplus$} (c23)
(c23) edge[red] node[labels] {$\ominus$} (l23)
(l23) edge[red] node[labels] {$\ominus$} (c24)
(c24.south west) edge[bend left=15,red] node[labels] {$\ominus$} (c21.south east)

(bi1) edge[dashed,bend left=10,red] node[labels,near end] {$\ominus$} (l21)
(l21) edge[dashed,red] node[labels,near start] {$\ominus$} (xi1)
(bi2) edge[dashed] node[labels,near end] {$\oplus$} (l22)
(l22) edge[dashed] node[labels,near start] {$\oplus$} (xi2)
(xi3) edge[dashed,bend right=10,red] node[labels,near end] {$\ominus$} (l23)
(l23) edge[dashed,red] node[labels,near start] {$\ominus$} (bi3)
;

\node[second node,left of=xi1,xshift=-1cm] (b) {$a_1$};
\node[main node,below of=b,yshift=-2cm] (a) {$a_k$};
\node[above right of=b] (invb1) {};
\node[right of=b] (invb2) {};
\node[below right of=b] (invb3) {};

\node[above right of=a] (inva1) {};
\node[right of=a] (inva2) {};
\node[below right of=a] (inva3) {};

\path[arcos]
(b) edge node[labels] {$\oplus$}  (a)
(invb1) edge[dashed] node[labels] {$\oplus$} (b)
(invb2) edge[dashed] node[labels] {$\oplus$} (b)
(invb3) edge[dashed] node[labels] {$\oplus$} (b)
(a) edge[dashed,red] node[labels] {$\ominus$} (inva1)
(a) edge[dashed,red] node[labels] {$\ominus$} (inva2)
(a) edge[dashed,red] node[labels] {$\ominus$} (inva3)
;

\end{tikzpicture}

%% file: fig/reduction_3.tex
\begin{tikzpicture}[node distance=1.75cm]
\setlength{\radio}{5cm}

\node[second node] (xi1) {$x_1$};
\node[main node,below right of=xi1] (fi1) {$f_1$};
\node[main node,above right of=xi1] (ti1) {$t_1$};
\node[second node,below right of=ti1] (bi1) {$\bar{x_1}$};
\node[main node] (t12) at ($(ti1)!0.5!(bi1)$) {};
\node[main node] (t13) at ($(ti1)!0.5!(fi1)$) {};
\node[main node] (f12) at ($(fi1)!0.5!(xi1)$) {};
\path[arcos]
(bi1) edge (fi1)
(fi1) edge (f12)
(f12) edge (xi1)
(xi1) edge (ti1)
(ti1) edge (t12)
(t12) edge (bi1)
(ti1) edge (t13)
(t13) edge (fi1)
;

\node[second node,right of=bi1] (xi2) {$x_2$};
\node[main node,below right of=xi2] (fi2) {$f_2$};
\node[main node,above right of=xi2] (ti2) {$t_2$};
\node[second node,below right of=ti2] (bi2) {$\bar{x_2}$};
\node[main node] (t22) at ($(ti2)!0.5!(bi2)$) {};
\node[main node] (t23) at ($(ti2)!0.5!(fi2)$) {};
\node[main node] (f22) at ($(fi2)!0.5!(xi2)$) {};
\path[arcos]
(bi2) edge (fi2)
(fi2) edge (f22)
(f22) edge (xi2)
(xi2) edge (ti2)
(ti2) edge (t22)
(t22) edge (bi2)
(ti2) edge (t23)
(t23) edge (fi2)
;

\node[second node,right of=bi2] (xi3) {$x_3$};
\node[main node,below right of=xi3] (fi3) {$f_3$};
\node[main node,above right of=xi3] (ti3) {$t_3$};
\node[second node,below right of=ti3] (bi3) {$\bar{x_3}$};
\node[main node] (t32) at ($(ti3)!0.5!(bi3)$) {};
\node[main node] (t33) at ($(ti3)!0.5!(fi3)$) {};
\node[main node] (f32) at ($(fi3)!0.5!(xi3)$) {};
\path[arcos]
(bi3) edge (fi3)
(fi3) edge (f32)
(f32) edge (xi3)
(xi3) edge (ti3)
(ti3) edge (t32)
(t32) edge (bi3)
(ti3) edge (t33)
(t33) edge (fi3)
;

\node[second node,below of=xi1,yshift=-2cm] (c11) {$c_{1,1}$};
\node[main node,right of=c11] (l11) {$\ell_{1,1}$};
\node[second node,right of=l11,xshift=0.25cm] (c12) {$c_{1,2}$};
\node[main node,right of=c12,xshift=-0.25cm] (l12) {$\ell_{1,2}$};
\node[second node,right of=l12,xshift=0.25cm] (c13) {$c_{1,3}$};
\node[main node,right of=c13] (l13) {$\ell_{1,3}$};
\node[second node,right of=l13,xshift=0.25cm] (c14) {$c_{1,4}$};

\node[main node] (s11) at ($(l11)!0.5!(c12)$) {};
\node[main node] (s12) at ($(l12)!0.5!(c13)$) {};
\node[main node] (s13) at ($(l13)!0.5!(c14)$) {};

\node[main node] (r11) at ($(l11)!0.5!(bi1)$) {};
\node[main node] (r12) at ($(l12)!0.5!(bi2)$) {};
\node[main node] (r13) at ($(l13)!0.5!(bi3)$) {};

\path[arcos]
(c11) edge (l11)
(l11) edge (s11)
(s11) edge (c12)
(c12) edge (l12)
(l12) edge (s12)
(s12) edge (c13)
(c13) edge (l13)
(l13) edge (s13)
(s13) edge (c14)
(c14.south west) edge[bend left=15] (c11.south east)

(xi1) edge[dashed] (l11)
(l11) edge[dashed] (r11)
(r11) edge[dashed] (bi1)
(xi2) edge[dashed] (l12)
(l12) edge[dashed] (r12)
(r12) edge[dashed] (bi2)
(xi3) edge[dashed] (l13)
(l13) edge[dashed] (r13)
(r13) edge[dashed] (bi3)
;

\node[second node,below of=c11,xshift=-0.5cm,yshift=-2cm] (c21) {$c_{2,1}$};
\node[main node,right of=c21] (l21) {$\ell_{2,1}$};
\node[second node,right of=l21,xshift=0.25cm] (c22) {$c_{2,2}$};
\node[main node,right of=c22] (l22) {$\ell_{2,2}$};
\node[second node,right of=l22,,xshift=0.25cm] (c23) {$c_{2,3}$};
\node[main node,right of=c23] (l23) {$\ell_{2,3}$};
\node[second node,right of=l23,xshift=0.25cm] (c24) {$c_{2,4}$};

\node[main node] (s21) at ($(l21)!0.5!(c22)$) {};
\node[main node] (s22) at ($(l22)!0.5!(c23)$) {};
\node[main node] (s23) at ($(l23)!0.5!(c24)$) {};

\node[main node] (r21) at ($(l21)!0.25!(xi1)$) {};
\node[main node] (r22) at ($(l22)!0.25!(xi2)$) {};
\node[main node] (r23) at ($(l23)!0.25!(bi3)$) {};

\path[arcos]
(c21) edge (l21)
(l21) edge (s21)
(s21) edge (c22)
(c22) edge (l22)
(l22) edge (s22)
(s22) edge (c23)
(c23) edge (l23)
(l23) edge (s23)
(s23) edge (c24)
(c24.south west) edge[bend left=15] (c21.south east)

(bi1) edge[dashed,bend left=10] (l21)
(l21) edge[dashed] (r21)
(r21) edge[dashed] (xi1)
(bi2) edge[dashed] (l22)
(l22) edge[dashed] (r22)
(r22) edge[dashed] (xi2)
(xi3) edge[dashed,bend right=10] (l23)
(l23) edge[dashed] (r23)
(r23) edge[dashed] (bi3)
;

\node[main node,left of=xi1,xshift=-1cm] (b) {$a_1$};
\node[main node,below of=b,yshift=-2cm] (a) {$a_k$};
\node[main node] (a2) at ($(b)!0.5!(a)$) {};

\node[above right of=b] (invb1) {};
\node[right of=b] (invb2) {};
\node[below right of=b] (invb3) {};

\node[above right of=a] (inva1) {};
\node[right of=a] (inva2) {};
\node[below right of=a] (inva3) {};

\path[arcos]
(b) edge (a2)
(a2) edge (a)
(invb1) edge[dashed] (b)
(invb2) edge[dashed] (b)
(invb3) edge[dashed] (b)
(a) edge[dashed] (inva1)
(a) edge[dashed] (inva2)
(a) edge[dashed] (inva3)
;

\end{tikzpicture}